\documentclass[twocolumn]{aastex63}
\usepackage{amsmath}
\DeclareMathOperator\arctanh{arctanh}
\received{}
\revised{}
\accepted{}
\submitjournal{ApJ}

\shorttitle{Lessons from SN Refsdal II}
\shortauthors{Zitrin A.}
\begin{document}

\title{Lessons from the first multiply imaged supernova II: A parametric strong-lensing model for the galaxy cluster MACS J1149.5+2223}

\correspondingauthor{Adi Zitrin}
\email{adizitrin@gmail.com}

\author[0000-0002-0350-4488]{Adi Zitrin}
\affiliation{Physics Department,
Ben-Gurion University of the Negev, P.O. Box 653,
Be'er-Sheva 84105, Israel}



\begin{abstract}
We present a parametric, grid-based lens model for the galaxy cluster MACS J1149.5+2223, concentrating on the properties of the first multiply imaged supernova Refsdal. This model complements our updated light-traces-mass (LTM) strong-lensing model for this cluster, described in a companion paper, and is generated using the same pipeline but with a different parametrization. Together these two models probe different possible solutions in a relatively self-consistent manner and can be used to examine systematic uncertainties and relevant differences between the two parameterizations. We obtain reasonably similar (agreeing to within $\simeq1-3\sigma$, in most cases) time delays and magnification ratios, with respect to S1, from the two different methods, although the LTM predictions seem to be systematically shorter/smaller for some of the images. Most notably, the time delay [and 95\% CI] between the Einstein cross (in particular, image S1), and SX, the image that appeared about a year after the original discovery of the cross, differs substantially between the parametric method (326 [300 -- 359] days) and the LTM method (224 [198 -- 306] days), which seems to underestimates the true reappearance time. The cause for this systematic difference is unclear at present. We speculate on its possible origin and note that a refined measurement of SN Refsdal's properties should help to more strongly discriminate between the two solutions, and thus between the two descriptions for the intrinsic shape of the underlying matter distribution. We also discuss the implications of our results for the Hubble constant.
\end{abstract}



\section{Introduction}\label{sec:intro}

Strong-lensing (SL) by galaxies and clusters of galaxies has become an important astrophysical and cosmological tool \citep[e.g.,][]{Kneib2011review}. %Due to the dependence on the projected mass density, lensing enables weighing and mapping the distribution of matter in such massive celestial objects \citep[e.g.][]{Caminha2016RXJ}, and thanks to its magnification.

On galaxy scales, a relatively rare chance alignment between lensing galaxies and distant background sources results in multiple images of the source, and often, an approximate  Einstein ring \citep[e.g.,][]{Kochanek2001Ring,Bolton2006SLACS}. If the background, multiply imaged source is variable, a measurement of the time delay (TD) between different multiple images of the source becomes possible, allowing, in conjunction with a lens model, a measurement of the Hubble constant \citep[e.g.,][]{Refsdal1964MNRAS}. For example, several strongly lensed quasars have been uncovered to date, some of them monitored over a long time span \citep[e.g.,][]{Courbin2011,Tewes2013,Suyu2013measured} to derive measurements of the Hubble constant, reaching, for example, an updated value of $H_{0}=73.3^{+1.7}_{-1.8}$ \citep[][]{Wong2019H0}. This value is in agreement with recent Cepheid calibration of Type Ia SN hosts \citep[][cf. Tip of the Red Giant Branch calibration \citealt{Freedman2019H0}]{Riess2019}, but in tension with that derived by \emph{Planck} measurements of the Cosmic Microwave Background \citep{Planck2018params}.    

On galaxy-cluster scales, space imaging with \emph{Hubble} has revealed that most massive clusters show an abundance of multiple images of lensed background sources near their center \citep[e.g.,][]{Richard2010locuss20,Zitrin2014CLASH25,Cerny2018,Paterno-Mahler2018,Acebron2019RXC0032}, where the projected matter density is sufficiently high. Hand in hand, over the past 2-3 decades various methods for analyzing strong cluster lenses have been developed \citep[e.g.,][]{Kneib1996,Broadhurst2005a,Diego2005Nonparam,Liesenborgs2006,Halkola2006,Jullo2007Lenstool}. Understanding the systematic biases and differences between the various methods and between the different parametrizations, i.e., different representations of the underlying matter distribution, is crucial for advancing the resulting science from SL, and for understanding and reducing the errors on the outcome measurements. Among the science themes that will gain from an improved understanding of lensing systematics are, for example, the measurements of $H_{0}$ \citep[e.g.,][]{Diego2016refsdal,Grillo2018H0m1149} or of other cosmological parameters \citep[e.g.,][]{Jullo2010}, the mapping of the distribution of matter in the lens \cite[e.g.,][]{Limousin2010M1423,Oguri201238clusters,Jauzac2015A2744,Caminha2016RXJ,Monna2017a611,Kawamata2016modelsHFF} and understanding its relation to the light distribution \citep[e.g.][]{Williams2011A3827,Massey2015,Chen2020A3827}, or the construction of luminosity functions for high-redshift lensed objects, to probe the reionizaion of the universe \citep[e.g.,][]{Bradac2012highz,Coe2014FF,Ishigaki2018HFF,Atek2015HalfHFFLF,Oesch2015shear,Mcleod2016,Salmon2020HighzRelics}. 

Various aspects of the systematics between different models have been investigated in recent years
\citep[e.g.,][]{Rodney2015A2744SN,Johnson2016systematics,Meneghetti2016SIMSCOMP,Acebron2017Systematics,BirrerTreu2019}. However, much work remains and additional information is often needed to break some of the degeneracies inherent to a typical lensing analysis, such as the apparent degeneracy between the intrinsic ellipticity of matter and the contribution from external substructure or external shear \citep[e.g.][see also \citealt{Kovner1987,NarayanBartelmann1996,Keeton1997Shear,HolderSchechter2003Shear,Meylan2006lensing}]{Zitrin2014CLASH25}. TDs and absolute magnifications (or in lack thereof, relative magnifications), hold the key to breaking some of these degeneracies, especially, in models that only rely on multiple-image positions as constraints, as is commonly the case. 

About five years ago, the first resolved multiply imaged SN was detected \citep{Kelly2015Sci} in the field of the galaxy cluster MACS J1149.5+2223 (M1149 hereafter, \citealt{EbelingMacs12_2007}), as an Einstein cross around a cluster galaxy close to the Brightest Cluster Galaxy (BCG). The SN appeared 50 years after the original paper by \citet{Refsdal1964MNRAS}, who proposed to use multiply imaged SNe to constrain the Hubble constant, and was named \emph{SN Refsdal}. The spiral galaxy in which the SN exploded is itself lensed several times by the cluster \citep{Zitrin2009_macs11495,Smith2009M1149}, so that another image of the exploding SN was expected about a year post-discovery. This enabled useful blind tests of various lens modeling techniques. While most models broadly agreed with the actual reappearance, these tests   \citep{Kelly2016reappearance,Treu2016Refsdal,Rodney2016Refsdal} revealed some discrepancies and supplied invaluable input for lens modelers to examine their modeling schemes.

In an accompanying paper (Zitrin, A., submitted and posted online; Paper I hereafter), we present a revised Light-Traces-Mass (LTM) lens model for M1149 and list the expected properties of SN Refsdal, namely the TDs and magnification ratios between the different images.
This revised LTM model also contains a fix to a numerical artifact that we uncovered following the blind tests mentioned above. Nevertheless, it appears that the LTM TD predictions (see Paper I) are systematically lower than what most parametric techniques predicted for some of the images, and SX in particular (\citealt{Kelly2016reappearance,Treu2016Refsdal,Rodney2016Refsdal}, see also \citealt{Sharon2015Refsdal,Grillo2018H0m1149,Oguri2015Refsdal}). Our goal here is to construct our own \emph{parametric} model \citep[e.g.][]{Zitrin2013M0416}, to both add our own parametric prediction to the relevant list, and to enable a more direct comparison with the LTM parametrization.

Our new, grid-based parametric model for M1149 is constructed on the same grid as the LTM model and using the same general pipeline, but with a different representation of galaxies and dark matter. We examine the differences between these two models with respect to the properties of SN Refsdal, and briefly discuss whether more robust TDs and relative magnifications for Refsdal might help distinguish between the models and thus reveal important information about the intrinsic matter distribution (a more profound examination of this is planned for future work).

The paper is organized as follows: in \S \ref{s:code} we give a short overview of the modeling method, and a description of its implementation to M1149 is given in \S \ref{s:modeling}. The results are presented and discussed in \S \ref{s:results}, with an emphasis on the properties of SN Refsdal, what it may teach us about the underlying matter distribution, and possible implications for the Hubble constant. The work is concluded in \S \ref{s:summary}. Throughout this work we use a $\Lambda$CDM cosmology with $\Omega_{M}=0.3$, $\Omega_{M}=0.7$, and $H_{0}=70$ km/s/Mpc. Unless otherwise stated, errors are $1\sigma$, and we generally use AB magnitudes \citep{Oke1983ABandStandards}.

\section{The lensing code}\label{s:code}
The parametric modeling scheme we use here \citep{Zitrin2013M0416,Zitrin2014CLASH25}, consists of two main constituents: galaxies and dark matter halos. Each galaxy is generally modeled as a double pseudo isothermal elliptical mass distribution (dPIE\footnote{This is similar to what is sometimes referred to as a Pseudo Isothermal Elliptical Mass Distribution, PIEMD, and is essentially a combination of two PIEMDs}; see \citealt{Eliasdottir2007}), and scaled following common scaling relations \citep[e.g.][see below]{Jullo2007Lenstool,Monna2017a611,Oguri2015Refsdal}. The projected density is for each galaxy is implemented as:
\begin{equation}
\Sigma(R)   = \frac{\sigma_0^2}{2G}  \frac{r_{cut}^2}{(r_{cut}^2-r_c^2)}\left(\frac{1} {\sqrt{r_c^2+R^2}}- \frac{1}{\sqrt{r_{cut}^2+R^2}} \right),
\end{equation}
where $G$ the gravitational constant, $r_{c}$ is the galaxy's core radius, $r_{cut}$ the cut-off radius, and $\sigma_0$ its velocity dispersion. $R$ is the 2D (elliptical) radius, given as:
\begin{equation}
\label{ecc}
{R}^2=\frac{X^2}{(1+\epsilon)^2} + \frac{Y^2}{(1-\epsilon)^2} ,
\end{equation}
with X and Y being the spatial coordinates along the major and minor axis, respectively. The relation between these and the original x and y coordinates of the grid, for a galaxy $i$ at $(x_{i},y_{i})$ with a position angle $\phi$, is defined by:

%\begin{equation}
\begin{equation} \label{grid}
\begin{array}{l}
X = (x - x_{i})\cos(\phi)+(y - y_{i})\sin(\phi) , \\
Y = -(x - x_{i})\sin(\phi)+(y - y_{i})\cos(\phi) ,
\end{array} 
\end{equation}
where if x and y are in pixels, an additional factor is needed to translate the grid's pixel coordinates to physical units. The eccentricity, $\epsilon$, is defined as $\epsilon\equiv (a-b)/(a+b)$, where $a$ and $b$ are the semi-major and semi-minor axes, respectively \citep{Eliasdottir2007}. Another common definition is the ellipticity, defined as $e\equiv 1-b/a$. In practice, only key (i.e., massive, central) galaxies we generally model as elliptical, whereas the rest of the galaxies are modeled as circular for speed-up purposes. 

The relevant quantities for each galaxy are scaled with the galaxy's luminosity $L$, compared to the values of some reference $L^\star$ galaxy at the cluster's redshift, using the following scaling relation:

\begin{equation}\label{scaling} 
\begin{array}{l}
\sigma_0  =   \sigma_0^\star (\frac{L}{L^\star} )^\lambda\;,  \\ 
r_{c}  =  r_{c}^\star (\frac{L}{L^\star} )^\beta\;, \\ 
r_{cut}  =   r_{cut}^\star (\frac{L}{L^\star} )^\alpha\,  \\
\end{array} 
\end{equation}
where we typically fix $\lambda=0.25$; $\beta=0.5$; $\alpha=0.5$. \\

% It is common to input a reference value $L^\star$, fix $r_{cut}$ to some fiducial value (typically between 0 and 0.5 kpc) and leave $\sigma_0$ and $r_{cut}$ free parameters. The exponents $\alpha$, $\beta$ and $\lambda$ are also kept fixed usually \citep[e.g.,[]{Jullo2007Lenstool,Grillo2018H0m1149}, although we can also leave them as free parameters \citep{Monna2017a611} to be optimized by the model. For example, a combination of $\alpha = 0.5$, $\beta=0.5$ and $\lambda=0.25$ describes a constant M/L, whereas increasing $\alpha = 0.8$ would describe the fundamental plane \citep[][and references therein]{Jullo2007Lenstool}.

% The galaxies component depends on a few parameters, most of them we can choose to fix -- based on some prior knowledge or common scaling relations -- while other we can leave free to be optimized in the minimizaiton.

The second constituent of the model, the DM halos (typically one to a few per cluster), are modeled here as elliptical Navarro-Frenk-White (eNFW; \citealt{Navarro1996}) mass distributions \citep[e.g.,][]{JingSuto2000,WrightBRainerd2000NFW,Meneghetti2003b,Oguri2015Refsdal}. NFW halos follow a radial mass-density profile of the form:

\begin{equation}
    \rho(R)=\frac{\rho_{s}}{(R/r_{s})(1+R/r_{s})^2}
\end{equation}
with $r_{s}$ being some scale radius, and $\rho_{s}$ the characteristic density of the halo. 

NFW halos can also be described by their concentration and mass. We mark by $\rho_{cr}$ the critical density of the universe at the redshift of the halo. The concentration we employ, $c_{200}$, is defined as $r_{200}$, the radius inside which the density of the halo equals 200$\rho_{cr}$, over the scale radius, $r_{s}$, such that $c_{200}=r_{200}/r_s$. The mass, $M_{200}$, is the mass enclosed within $r_{200}$. 

% The relation between the different NFW representations can be written as:

% \begin{equation}
% \left\{
% \begin{array}{l}
% \rho_{s}=\frac{200}{3}\rho_{cr}\frac{c_{200}^3}{[\ln(1+c_{200})-c_{200}/(1+c_{200})]} \\
% M_{200}=\frac{800\pi}{3} \rho_{cr} r_{200}^3 .
% \end{array}
% \right.
% \end{equation}
% \\

The projected eNFW surface mass-density distribution is implemented in our code using these concentration and mass definitions, as:

\begin{equation}
\Sigma(r) = \frac{2\rho_s r_{s}}{r^{2}-1}f(r) ,
\end{equation}
with

\begin{equation}
\rho_{s}=\frac{200}{3}\rho_{cr}\frac{c_{200}^3}{ [\ln(1+c_{200})-c_{200}/(1+c_{200}) ] } ,
\end{equation}

\begin{equation}
r_{s}=\left(\frac{1}{\pi} \frac{3}{800} \frac{1}{\rho_{cr}} \frac{M_{200}}{c_{200}^3}\right)^\frac{1}{3} ,
\end{equation}
and

\begin{equation}
f(r)=    \left\{ \begin{array}{l}
    1-\frac{2}{\sqrt{r^{2}-1}}\arctan\sqrt{\frac{r-1}{r+1}}~~~~~ (r>1) \\
    1-\frac{2}{\sqrt{1-r^{2}}}\arctanh\sqrt{\frac{1-r}{1+r}}~~~~ (r<1) \\
    ~~~~~~~~~~~~~~~~~~~~~~~~~0~~~~~~~~~ (r=1) , \\
    \end{array}
   \right.
\end{equation}
where the dimensionless $r$ is the 2D (elliptical) radius in units of $r_{s}$, i.e., $r\equiv R/r_{s}$, and the ellipticity and translation to the grid's coordinates are introduced as above (eqs. \ref{ecc} \& \ref{grid}).

While analytic expressions for the potential and the deflection fields exist, these are obtained in our procedure by integrating (in Fourier space) the mass density map, where the relation between the (surface) mass-density map obtained above and the effective lensing potential is given by  \citep{NarayanBartelmann1996}:
\begin{equation}
\Psi(\vec\theta)=\frac{1}{\pi}\int \kappa(\vec\theta') \ln |\vec\theta-\vec\theta'| d^2\theta' ,
\end{equation}
where $\kappa$ is the surface mass density in units of the critical density for lensing, i.e., $\kappa=\Sigma/\Sigma_{crit}$, and $\Sigma_{crit}=\frac{c^2}{4\pi G}\frac{D_{s}}{D_{ls}D_{l}}$, with $c$ the speed of light, and $D_{l}$, $D_s$, $D_{ls}$ the angular diameter distances to the lens, to the source, and between the lens and the source, respectively.

The relation between the surface mass-density map and the deflection field is given by \citep{NarayanBartelmann1996}:
\begin{equation}
\vec\alpha(\vec\theta)=\nabla\Psi=\frac{1}{\pi}\int \kappa(\vec\theta')\frac{\vec\theta-\vec\theta'}{|\vec\theta-\vec\theta'|^2} d^2\theta' .
\end{equation}
\\

To avoid numerical artifacts described in Paper I, in each step the potential is calculated as above and constitutes a new starting-point for the model from which all lensing quantities are then self-consistently derived. 

In addition to the two main components, i.e., the galaxies and dark matter halos, an external shear can be in principle added directly to the deflection field. 

We often leave as free parameters the core size, relative weight (i.e., the mass or relative mass-to-light ratio), ellipticites and position angles of key cluster members, such as the BCGs. The optimization of the model is performed with the same pipeline as for our LTM model (Paper I), using a $\chi^2$ function that minimizes the distance between the predicted multiple images (using a simply averaged source position for each system) and their observed locations. The minimization is carried out with a Monte-Carlo Markov Chain with a Metropolis-Hastings algorithm. Punishing terms can be added for images with wrong parity, or if extra images are predicted. Some annealing is also included in the process, and the chain typically runs for several thousand steps after the burn-in phase. Errors are then typically calculated from the same Markov chain.

We refer to the model, in short, after the combination of underlying forms adopted for the galaxies and dark matter, i.e., dPIEeNFW.

\begin{figure*}
 \begin{center}
 %\vspace{0.2cm}
  \includegraphics[width=0.49\textwidth,trim=0cm 0cm 0cm 0cm,clip]{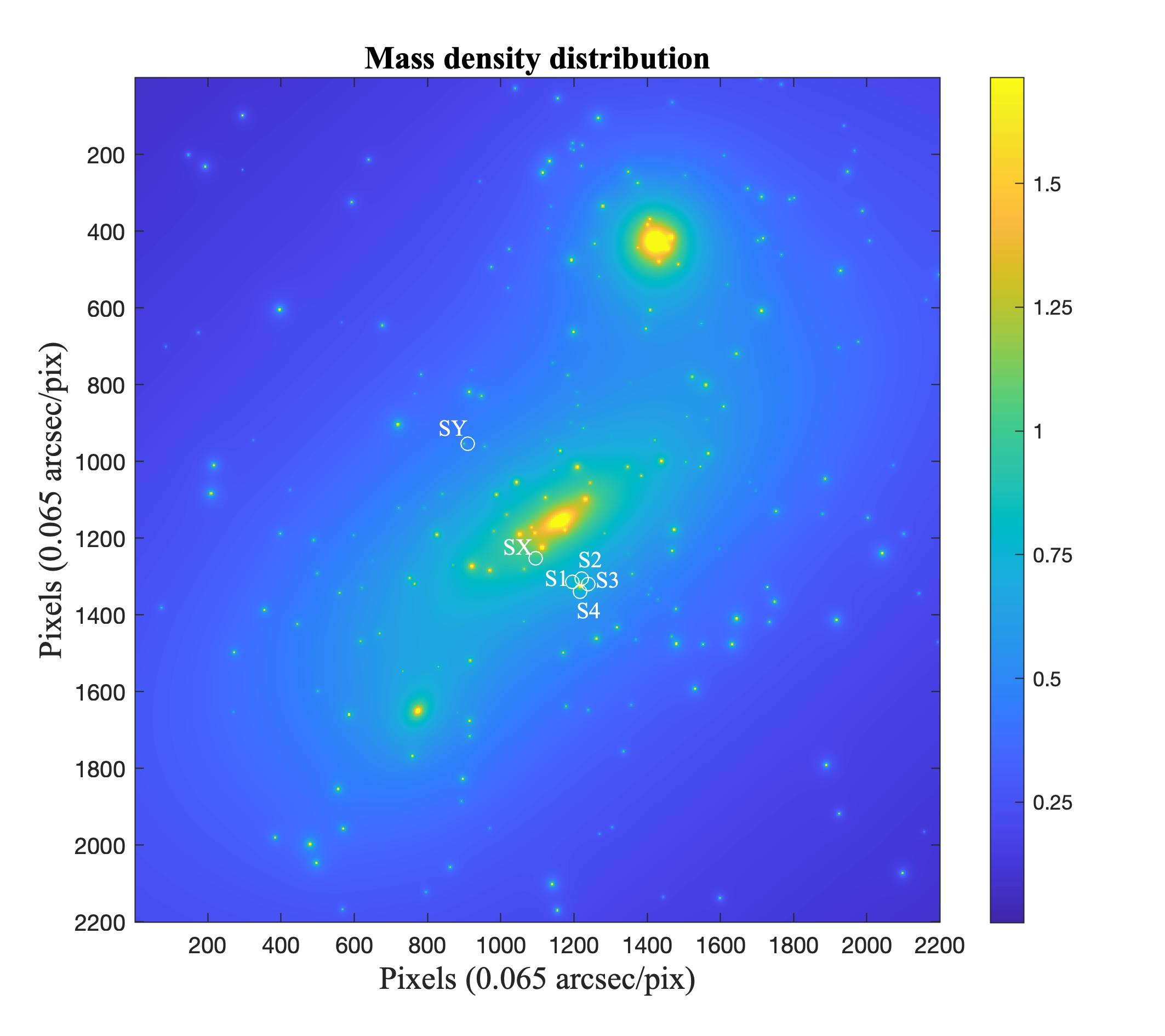}
    \includegraphics[width=0.49\textwidth,trim=0cm 0cm 0cm 0cm,clip]{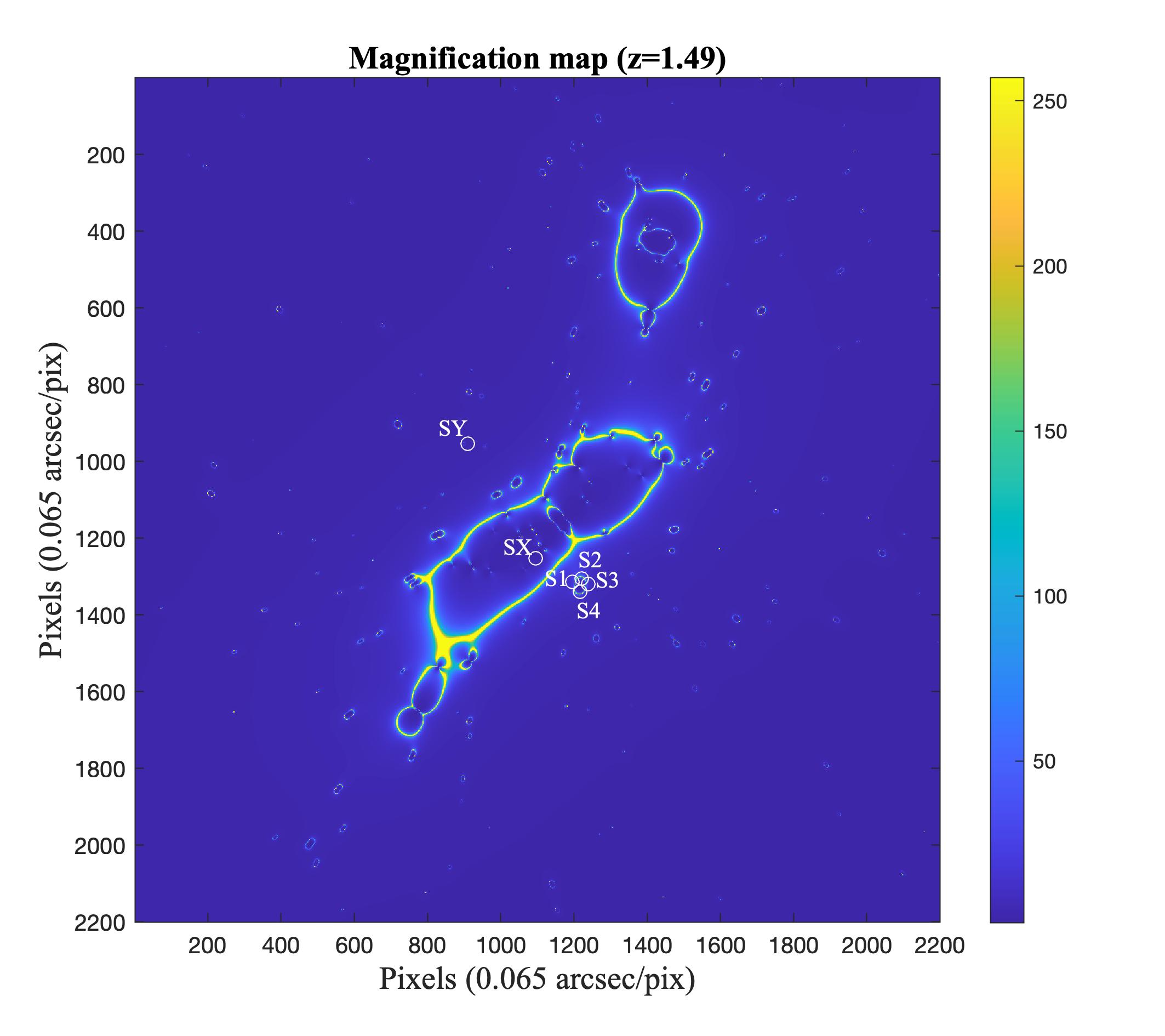}\\
      \includegraphics[width=0.49\textwidth,trim=0cm 0cm 0cm 0cm,clip]{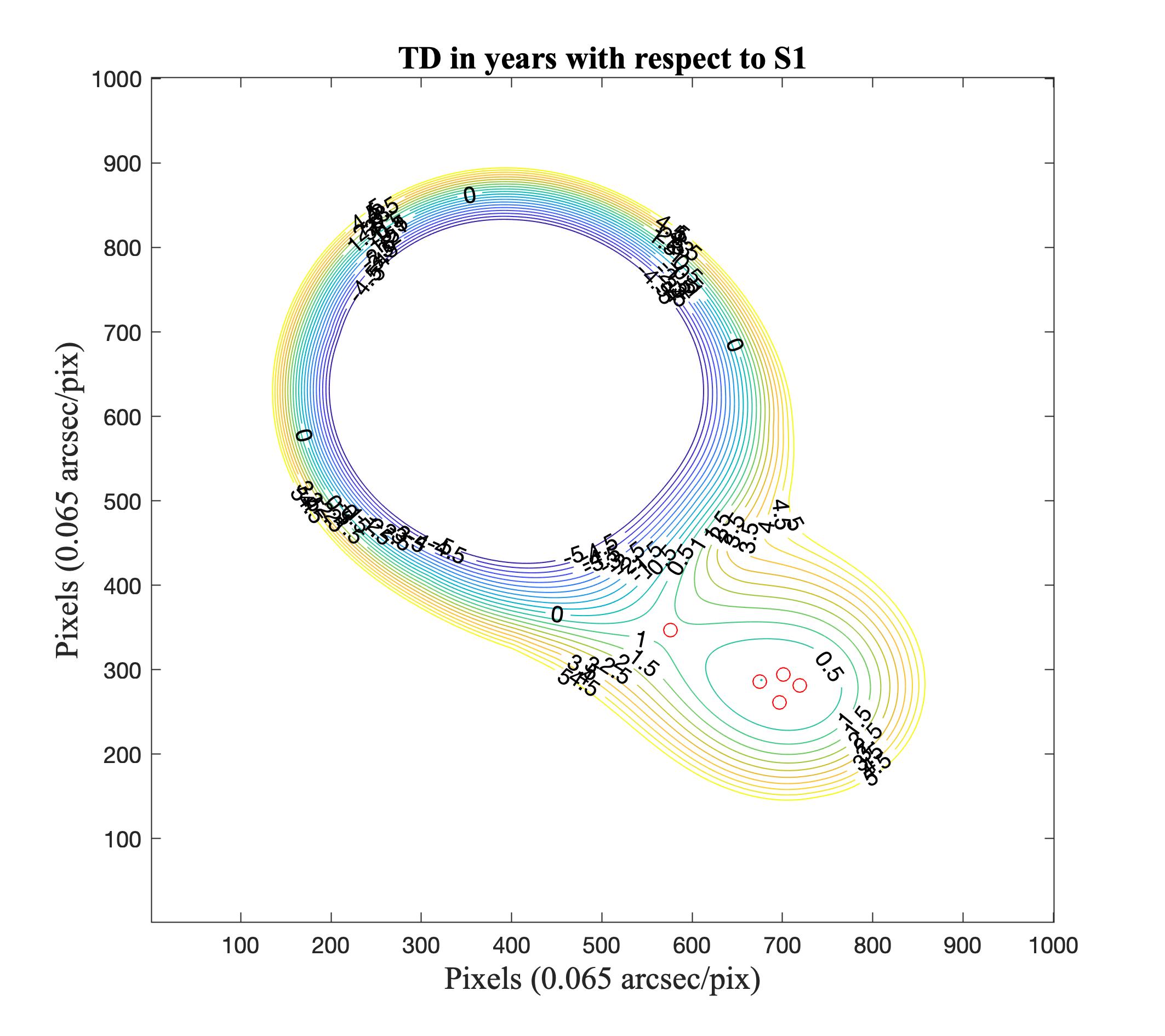}
    \includegraphics[width=0.49\textwidth,trim=0cm 0cm 0cm 0cm,clip]{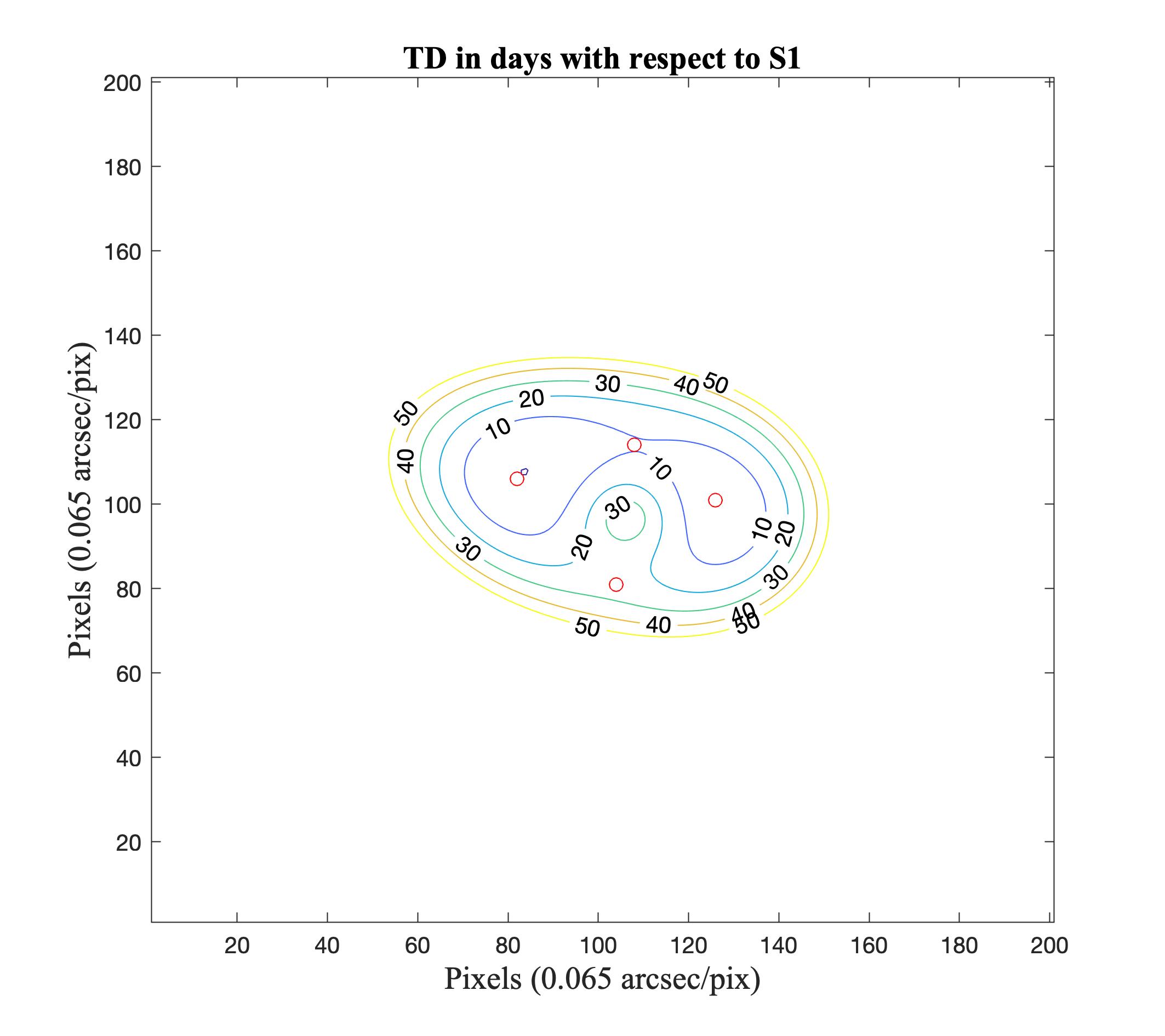}
 \end{center}
\caption{Our parametric dPIEeNFW model for M1149. \emph{Upper left} figure shows $\kappa$, the surface mass density in units of the critical density for lensing, scaled to the redshift of system 1, the spiral galaxy hosting SN Refsdal, at $z=1.49$. The positions of S1-S4, SX, and SY are marked with white circles. \emph{Upper right} figure shows the magnification map from the model for that redshift, similarly marking the SN image positions. \emph{Bottom left} figure shows contours of the TD surface with respect to the SN image S1, in decrements of 0.5 years, marking with red circles the Einstein cross images and and SX. \emph{Bottom right} figure shows contours of the TD surface with respect to the SN image S1, in decrements of 10 days, marking with red circles the positions of the Einstein cross images.}\label{fig:kappa_mag_TD}
\end{figure*}

% The LTM code, presented in more detail in Paper I (see also \citealt{Zitrin2009_cl0024,Zitrin2014CLASH25}), and to which we compare our results, relies on the assumption that light traces mass, so that the weighted luminosity distribution of cluster galaxies can act as a guide for the shape of the total matter distribution, including dark matter. The code is run via a long MCMC chain after burning, on an grid which is the same as Hubble original images, or a few times lower for speed up purposes. The same pipelines also include a parametric implementation, which we use here and now detail. 

\section{\lowercase{d}PIE\lowercase{e}NFW modeling of M1149}\label{s:modeling}
For constraining the model for M1149 we use the same list of constraints as in Paper I, namely the \emph{gold} sample from \citet{Treu2016Refsdal} and \citet{Finney2018M1149}, with some minor additions of \emph{silver} images. Images were vetted and ranked by lens modelers in efforts surrounding the Hubble Frontier Fields program \citep{Lotz2017HFF}, with the gold images obtaining the highest scores and rendered as most secure (\citealt{Treu2016Refsdal} for more details). We also include the list of knots for the spiral galaxy from \citet{Treu2016Refsdal} and \citet{Finney2018M1149}. To avoid unnecessary duplication, we refer the reader to Paper I; the constraints are listed in Tables 2 \& 3 therein. Similar to the LTM model, we employ for the $\chi^2$ function a positional uncertainty of $\sigma_{pos}=0.5\arcsec$ for most images, although for the four Einstein cross SN images we adopt $\sigma_{pos}=0.1\arcsec$. Most of the systems have a spectroscopic measurement that was used to anchor their lensing distance. Systems 6 \& 7 do not have a spectroscopic measurement available and we fix their redshift to $\simeq2.6$ based on their photo-$z$. We do not use any measured TDs or magnification ratios as input. 

We include the same list of cluster galaxies as in our LTM model - namely a red-sequence based selection, cross-matched with available spectroscopic information \citep[][and references therein]{Treu2016Refsdal}. We fix the ellipticity of the three brightest galaxies (RA=11:49:35.70, Dec=+22:23:54.71; RA=11:49:36.9328, Dec=+22:25:35.882; RA=11:49:37.55, Dec=+22:23:22.49) to their measured values and refer to all other galaxies as circular. We employ a vanishing core for all galaxies, i.e., fixing $r_{core}^\star=0$, but allow for a finite core for the BCG, leaving the core size free to be optimized in the minimization.%, and for the second and third brightest galaxies we scale it with $r_c^{*}=0.2$ kpc. %We leave free the scaling relation exponent $\alpha$ and $\lambda$, responsible for the scaling of the velocity dispersion, and cut-off radius, respectively.

Following previous work \citep[e.g.,][]{Oguri2015Refsdal, Grillo2018H0m1149,Sharon2015Refsdal,Kawamata2016modelsHFF}, we include in the modeling three dark matter halos: one is centered on the BCG (RA=11:49:35.70, DEC=+22:23:54.71), with the exact position to be optimized in the modeling; one centered on a bright galaxy about 40\arcsec\ south-east of the BCG (RA=11:49:37.55, DEC=+22:23:22.49), and one around a clump of galaxies about 50\arcsec\ roughly north of the BCG (at RA=11:49:34.52, DEC=+22:24:42.09), whose exact position is also free to be determined in the minimization. We do not include an external shear in the current model.

We build our model on the same grid as the LTM model (Paper I), concentrated on the central $\sim140\arcsec \times140\arcsec$ area of the cluster, and with a resolution of 0."065/pix, native to the public CLASH HST images (some parts in the minimization are done in a few-times lower resolution for speed-up purposes, but the final chain is run with the native resolution).

\begin{figure*}
 \begin{center}
 %\vspace{0.2cm}
  \includegraphics[width=0.95\textwidth,trim=0cm 0cm 0cm 0cm,clip]{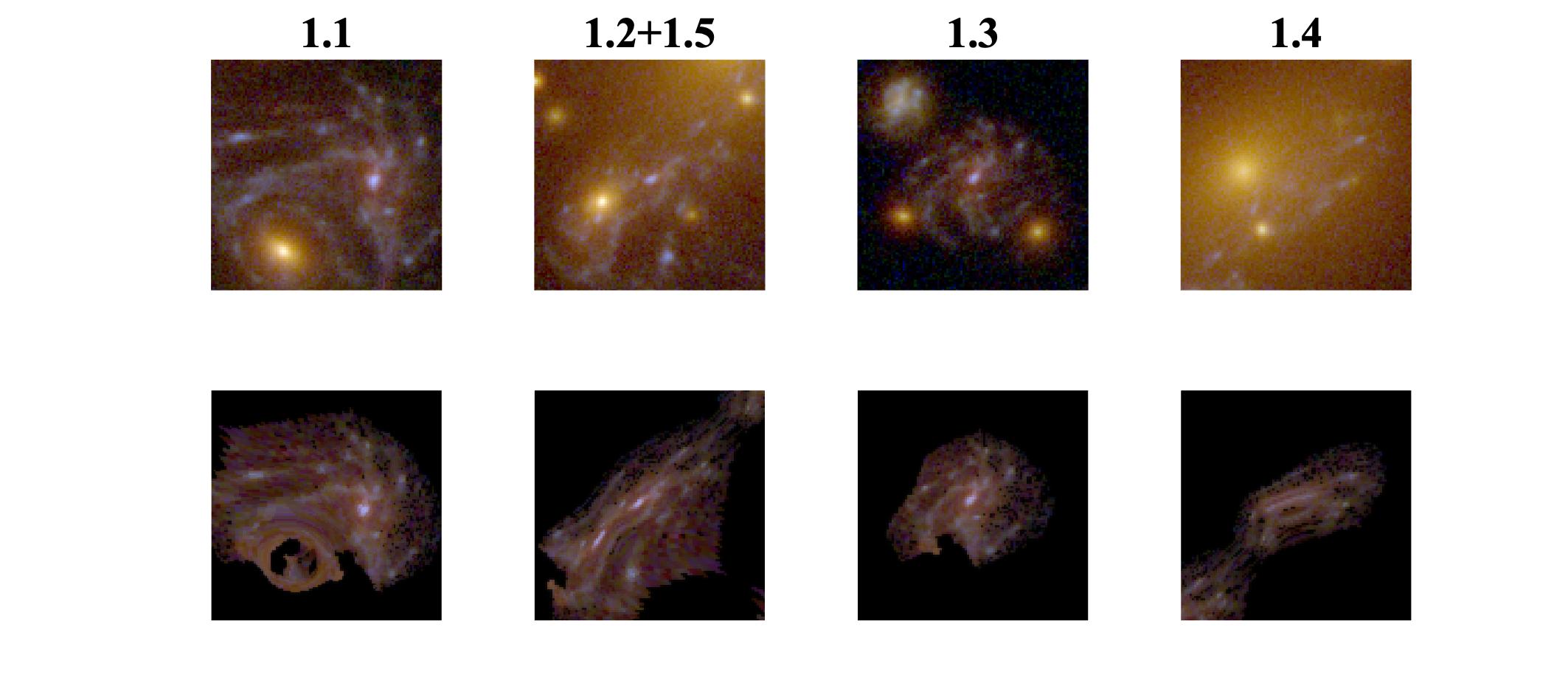}
 \end{center}
\caption{Reproduction of system 1, the spiral galaxy hosting SN Refsdal, by our parametric model. We send the largest image, 1.1, to the source plane and back, to obtain the reproduction of the other images of this system. The first stamp is 7.8\arcsec on 7.8\arcsec and the three other stamps are 6.5\arcsec on 6.5\arcsec.}\label{fig:Sys1Spiral}
\end{figure*}

\section{Results}\label{s:results}

We show our resulting surface mass-density and magnification maps, as well as TD surface contours, in Fig. \ref{fig:kappa_mag_TD}. The final model has an image reproduction \emph{rms} of 0.57\arcsec, and a $\chi^2\simeq170$. The number of constraints following the multiple-image list is $N_{c}=170$, and the number of free parameters is $N_{p}=21$. Correspondingly, the number of degrees of freedom is DOF$=N_{c}-N_{p}=149$. The \emph{reduced} $\chi^2$ is thus $\simeq1.14$. For comparison, the LTM has an \emph{rms} of 0.68\arcsec, a $\chi^2\simeq242$, and a reduced $\chi^2\simeq1.5$, such that both are of roughly similar accuracy (the \emph{rms} is $\sim15\%$ lower for the parametric model) but the parametric method does seem to yield a better fit overall and better match to the data.

As another indication of the model's accuracy, we also show here (Fig. \ref{fig:Sys1Spiral}; as was also done for our LTM model in Paper I), the reproduction of system 1, the spiral galaxy, by the model --  compared to the data. We delens its largest image to the source plane and back through the lens to form the other images of the system. Although some local inaccuracies exist, the detailed reproduction is evidently very successful. 

\begin{deluxetable*}{lllll}
\tablecaption{Time Delays and Magnification ratios for SN Refsdal}
\label{TDs}
\tablecolumns{5}
\tablewidth{0.9\linewidth}
\tablehead{
\colhead{Parameter
} &
\colhead{$\Delta t(t)$
} &
\colhead{$\Delta t(p)$
} &
\colhead{LTM [68.3\% CI] [95\% CI]
} &
\colhead{dPIEeNFW [68.3\% CI] [95\% CI]
}
}
\startdata
$\Delta t_{S2:S1}$ & $4 \pm 4^{a}$ & $7 \pm 2^{a}$ & 5.4 [3.3 -- 6.4] [2.8 -- 7.2]& 9.8 [9.6 -- 11.8]  [9.0 -- 12.8] \\
$\Delta t_{S3:S1}$& $2 \pm 5^{a} $&$0.6 \pm 3^{a}$& 1.6 [0.8 -- 2.1] [0.6 -- 2.9] & 3.0 [2.9 -- 4.0]  [2.7 -- 4.7]\\
$\Delta t_{S4:S1}$& $24 \pm 7^{a}$& $27 \pm 8^{a}$& 26.3 [23.4 -- 27.4] [22.6 -- 28.3]& 24.0 [23.6 -- 25.8]  [22.8 -- 27.0]\\
$\Delta t_{SX:S1}$& $345 \pm 10^{b}$& $345 \pm 10^{b}$ & 224.4$^{\dagger}$ [221.4 -- 272.7] [197.8 -- 305.5]  & 325.8 [311.6 -- 344.6]  [299.6 -- 358.5]\\
$\Delta t_{SY:S1}$& ---&  --- & -6522 [-6623 -- -6137] [-6834 -- -6025]& -6418 [-6358 -- -6022]  [-6735 -- -5832]\\
$\mu_{S2}/\mu_{S1}$& $1.15 \pm 0.05^{a}$ & $1.17 \pm 0.02^{a}$& 0.86  [0.69 -- 1.34] [0.45 -- 1.57]& 1.17 [0.84 -- 1.13]  [0.75 -- 1.30]\\
$\mu_{S3}/\mu_{S1}$& $ 1.01 \pm 0.04^{a}$ & $1.00 \pm 0.01^{a} $& 0.94 [0.88 -- 1.00] [0.78 -- 1.09]& 1.13 [1.13 -- 1.26]  [1.12 -- 1.33] \\
$\mu_{S4}/\mu_{S1}$& $ 0.34 \pm 0.02^{a} $ & $0.38 \pm 0.02^{a}$& 0.23 [0.19 -- 0.32] [0.14 -- 0.36]& 0.66 [0.52 -- 0.66] [0.44 -- 0.72] \\ 
$\mu_{SX}/\mu_{S1}$& $0.28\pm0.1^{b}$& $0.28\pm0.1^{b}$ &0.21 [0.19 --  0.23] [0.16 -- 0.25] & 0.25 [0.24 -- 0.26]  [0.23 -- 0.27]\\
$\mu_{SY}/\mu_{S1}$ & --- & --- & 0.15 [0.13 -- 0.17] [0.11 -- 0.18]& 0.24 [0.22 -- 0.24]  [0.21 -- 0.25]\\
\enddata
\tablecomments{Early measurements of SN Refsdal's TDs (in days) and magnification ratios, along with estimates from our new LTM model. Note that TDs and magnification ratios were not used as constraints in the minimization, and similarly, any information regarding SX (and SY) was not used explicitly as well. We anticipate updated and more accurate measurements for SN Refsdal will become available in the future.\\
$^{a}$ - Taken from Table 3 in \citet{Rodney2016Refsdal} based on a set of templates (t), or on polynomials (p) [this notation is adopted from \cite{Grillo2018H0m1149}].\\ 
$^{b}$ - Taken from \cite{Grillo2018H0m1149}, as estimated from Fig. 3 in \citet{Kelly2016reappearance}.\\ 
$^{\dagger}$ - If instead the source position is derived, in addition to the Einstein cross images, using also the position of SX, the LTM TD about 40-50 days longer and likely closer to the true value (see Paper I).\\
}
\end{deluxetable*}

Our model is made publicly available \footnote{\url{https://www.dropbox.com/sh/xkzysm4nhyq28e3/AABJVCWVzbkGdnvyD67CODyha?dl=0}}. 

\subsection{dPIEeNFW TDs and magnifications for SN Refsdal}

Our main goal here is to obtain estimates for the TDs and magnification ratios for SN Refsdal from our dPIEeNFW model, so these could be later compared with more accurate measurements for the SN and, with our LTM model (Paper I), as well as with other models that supplied estimates for the SN \citep[][and references therein]{Treu2016Refsdal,Kelly2016reappearance,Rodney2016Refsdal}. 
The model TDs with respect to S1 are given in Table \ref{TDs}. The quoted errors were derived using a 100 random models from a designated MC chain, and we list therein both the 68.3\% and 95\% confidence intervals. These were obtained by adopting an effective $\sigma_{pos}\simeq0.7\arcsec$, which better encompasses the range of values from different dPIEeNFW models we generated during the final modeling of M1149. 

Our model predicts that among the Einstein cross images, S1 arrived first; S3 arrived second, about 3 days later; S2 arrived about 10 days after S1; and S4 about $\simeq$24 days after S1 (see Table \ref{TDs} for exact numbers). The SX-S1 TD and 95\% CI is predicted to be $\sim326$ [300 -- 359] days. According to our model, SY appeared close to 18 years ago. The best-fit magnifications we obtain for the different images are [17.3, 20.2, 19.6, 11.4, 4.4] for [S1, S2, S3, S4, SX, SY], respectively. The magnification ratios and their uncertainties are listed in Table \ref{TDs} as well.

\begin{figure*}
 \begin{center}
  \includegraphics[width=0.98\textwidth,trim=5cm 2cm 5cm 2cm,clip]{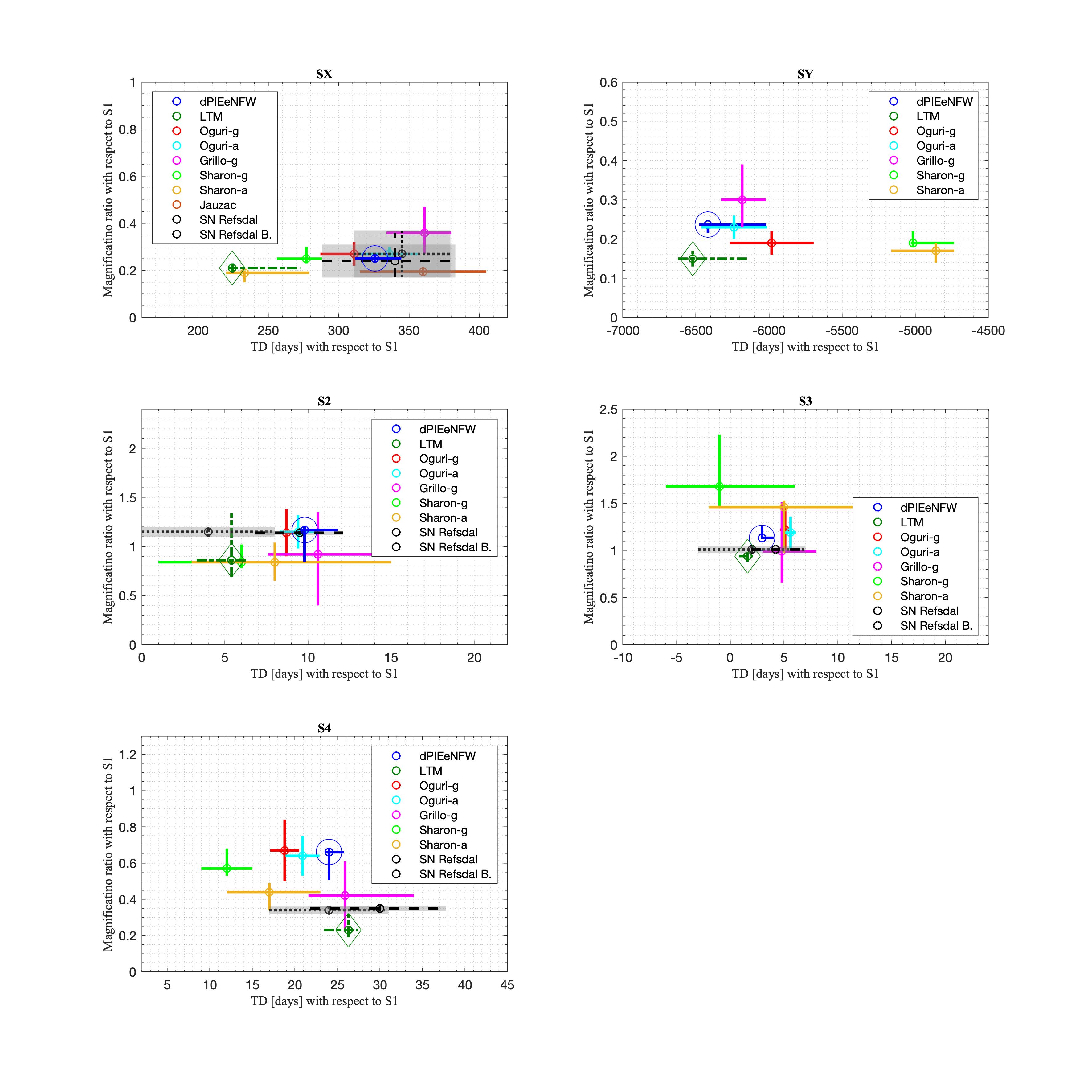}
 \end{center}
\caption{Comparison of TDs and magnification ratios from our parametric dPIEeNFW model and our new LTM model with early measurements of SN Refsdal and other lens models. All measurements are with respect to S1. Our dPIEeNFW model is plotted in \emph{blue} and marked for emphasis with a \emph{larger circle}. Our LTM model is plotted with a \emph{dash-dotted darker green line} and marked with a \emph{larger diamond} for emphasis. We also overplot with a \emph{black dotted line} embedded in a \emph{grey rectangle} the early measurements for SN Refsdal, where available (SY does not have a measurement). These are read of Figure 3 in \cite{Kelly2016reappearance} for SX (showing here only approximate uncertainties and not the full confidence region shape shown therein), and from \citet{Rodney2016Refsdal} for S2, S3, S4 using their template-fit results. We also show in \emph{black dashed line} embedded in a \emph{grey rectangle}, the recent measurements from \cite{Baklanov2020}. Other \emph{parametric} lens models submitted to the blind test by \citet{Treu2016Refsdal} are also over-plotted. Most values were taken from Table 6 therein, whereas the \emph{Jauzac} entry was read off Fig. 3 in \citet{Kelly2016reappearance}.}\label{fig:compwithMeasurements}
\end{figure*}

In Fig. \ref{fig:compwithMeasurements} we plot our dPIEeNFW estimates for SN Refsdal, along with the early measurements  by \citet{Kelly2016reappearance} and \citet{Rodney2016Refsdal}. We also show therein, for comparison, predicitions from other parametric models that appeared in \citet{Treu2016Refsdal} as well as from our LTM model from Paper I. The values from our parametric, dPIEeNFW model seem to be in a very good agreement (typically to within 1 or 2 $\sigma$) with the predictions from the other parametric models applied to M1149, especially with those from the \emph{Grillo} and \emph{Oguri} models (see \citealt{Kelly2016reappearance} and Table 6 in \citealt{Treu2016Refsdal}). They also agree well with the early SN measurements shown therein and listed here in Table \ref{TDs} (although in some cases there is some disagreement -- for example for S4). In addition, during the writing of this work another measurement was published by \citet{Baklanov2020} by using the \citet{Kelly2016reappearance} photometry and hydrodynamics modeling of the explosion, which we also include in Fig. \ref{fig:compwithMeasurements}. These new measurements seem to be in very good agreement with the numbers from our parametric model as well. 

Despite the good agreement with both the predictions from other models \citep[see also][]{Oguri2015Refsdal,Grillo2018H0m1149}, and with the measurements of the observed SN images, it should be noted that the modeling scheme was not tailored to model the Einstein cross in high detail: First, we work on a grid, and so our solution is limited to the grid's finite resolution, which is non-negligible compared to the distances of the cross' images from the lensing galaxy's critical curves, for example. Second, while we leave free the weight (or mass) of the galaxy around which the Einstein cross forms, we include no ellipticity for it, nor do we separately optimize its core and cut-off radii (these instead were assumed to follow the scaling relation, where the core radius was set to zero here). We speculate that \emph{analytic} methods, i.e., those not confined to a certain grid resolution, could in principle obtain higher-resolution results where such accuracy is needed. In the case of SN Refsdal, however, it seems that despite these limitations our grid-based model is comparable to that from other parametric models.

\subsection{Comparison with the LTM model}\label{ss:older}
While our parametric model agrees very well with the typical predictions from other parametric lens models that were applied to SN Refsdal \citep[Fig. \ref{fig:compwithMeasurements}; see also][]{Kelly2016reappearance,Treu2016Refsdal}, the LTM model (Paper I) seems to yield somewhat smaller magnification ratios for all images, with respect to S1, and systematically shorter TDs for three out of the five examined TDs: SX-S1, S2-S1, and S3-S1. While most (although not all) parametric methods, including our own presented here, yield a TD of order $\sim300-360$ days for the appearance of SX after S1, the LTM model suggests a systematically lower TD 95\% CI of $\sim200-300$ days  (or $\sim$250-320 days, if adopting a revised source position using information on SX). For the Einstein cross images S2 and S3, the LTM  model suggests 6 days instead of 10 for the S2-S1 TD, and 1-2 days instead of 3-4 days for the S3-S1 TD. The S4-S1 TD seems to be similar in the two models.

\begin{figure*}
 \begin{center}
 %\vspace{0.2cm}
  \includegraphics[width=0.35\textwidth,trim=0cm 0cm 0cm 0cm,clip]{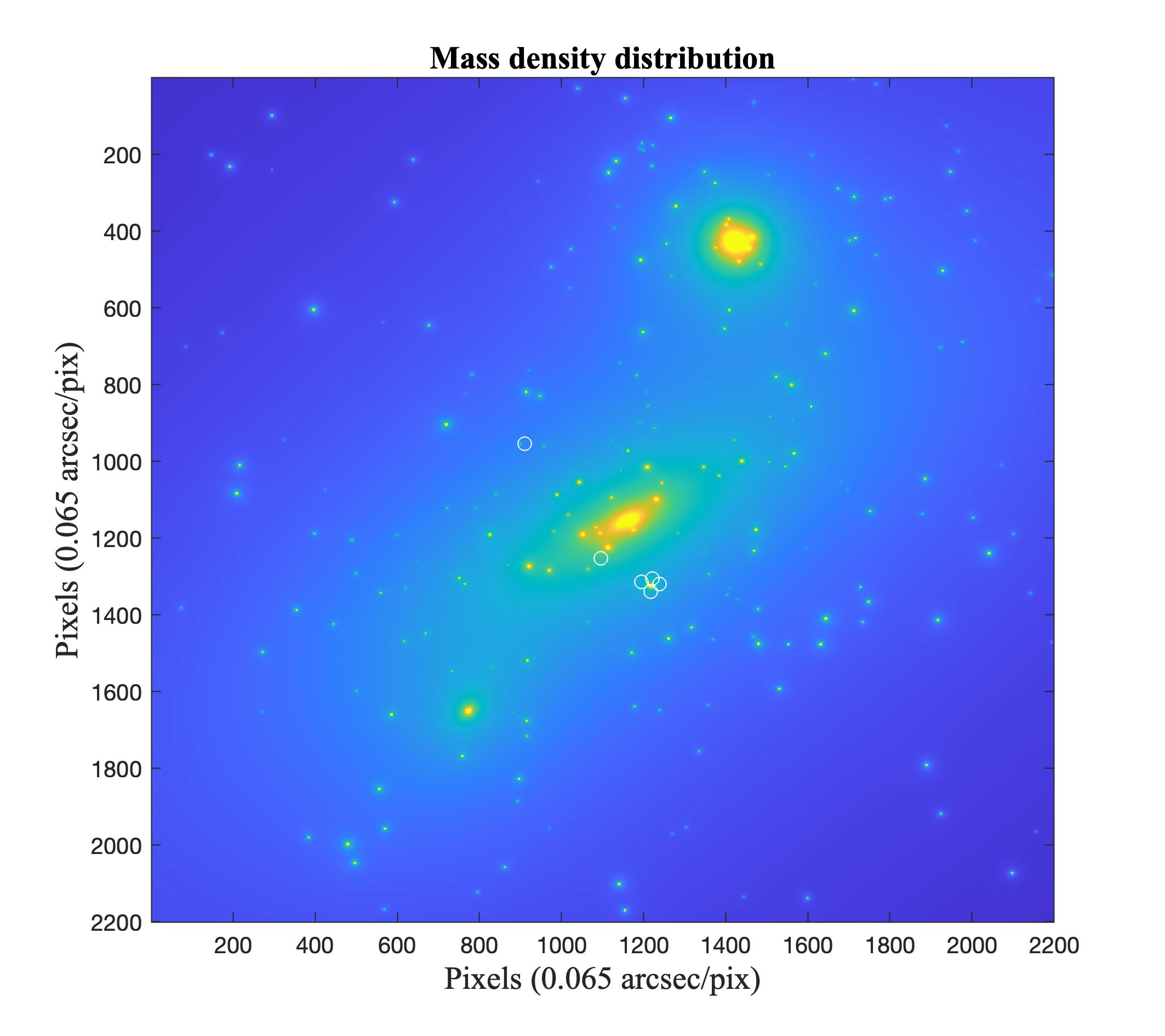}
    \includegraphics[width=0.35\textwidth,trim=0cm 0cm 0cm 0cm,clip]{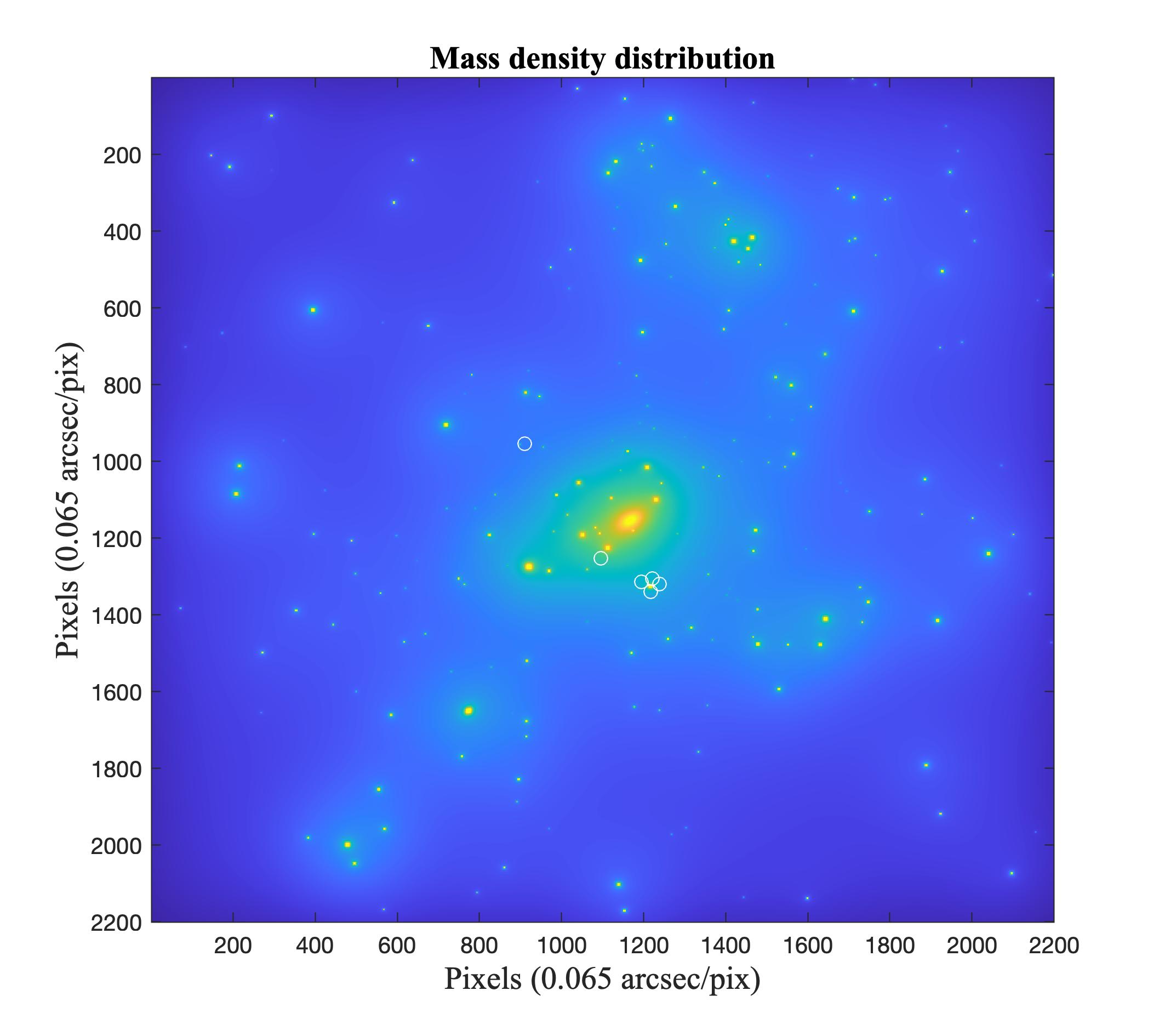}\\
    \includegraphics[width=0.35\textwidth,trim=0cm 0cm 0cm 0cm,clip]{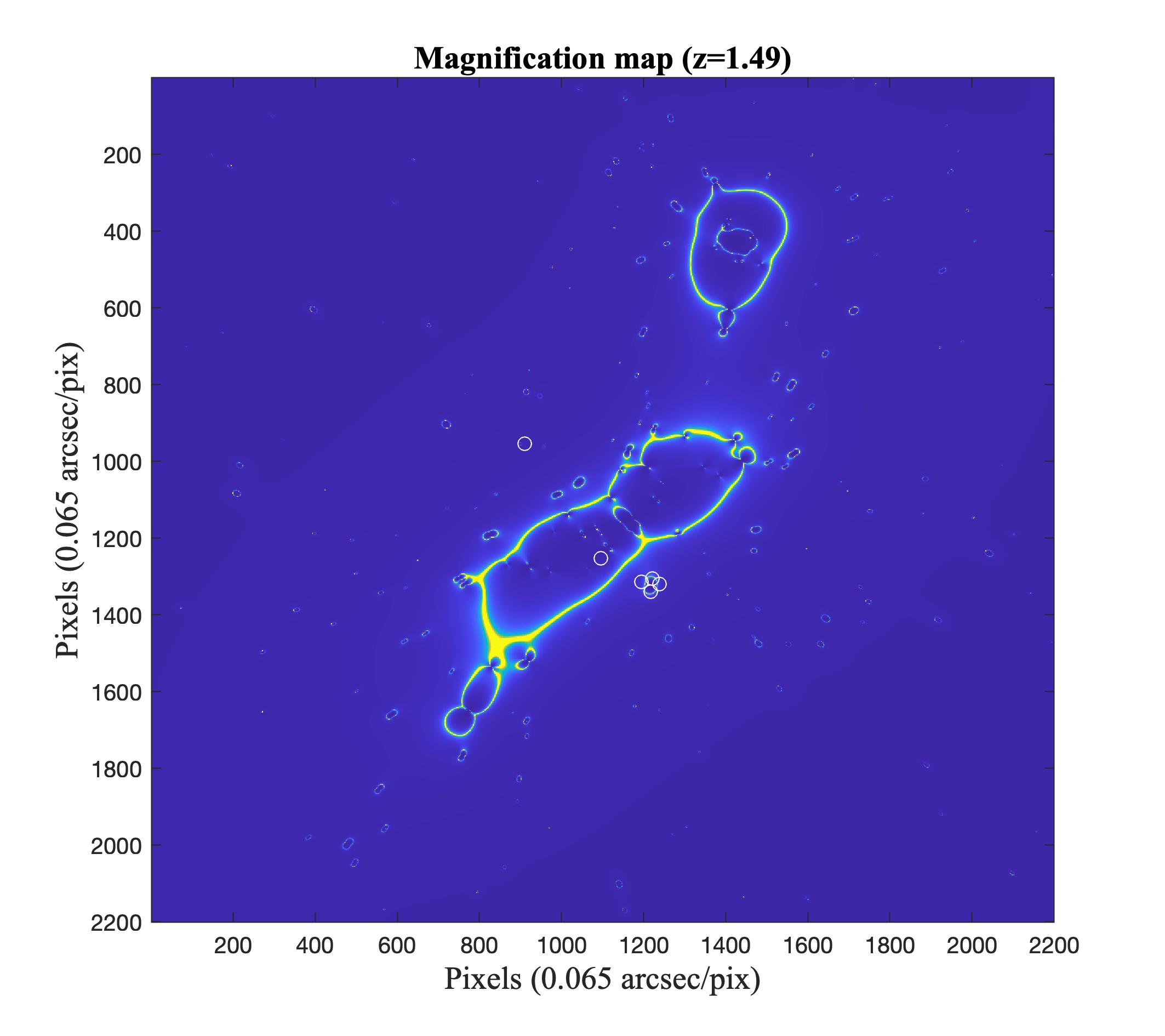}
    \includegraphics[width=0.35\textwidth,trim=0cm 0cm 0cm 0cm,clip]{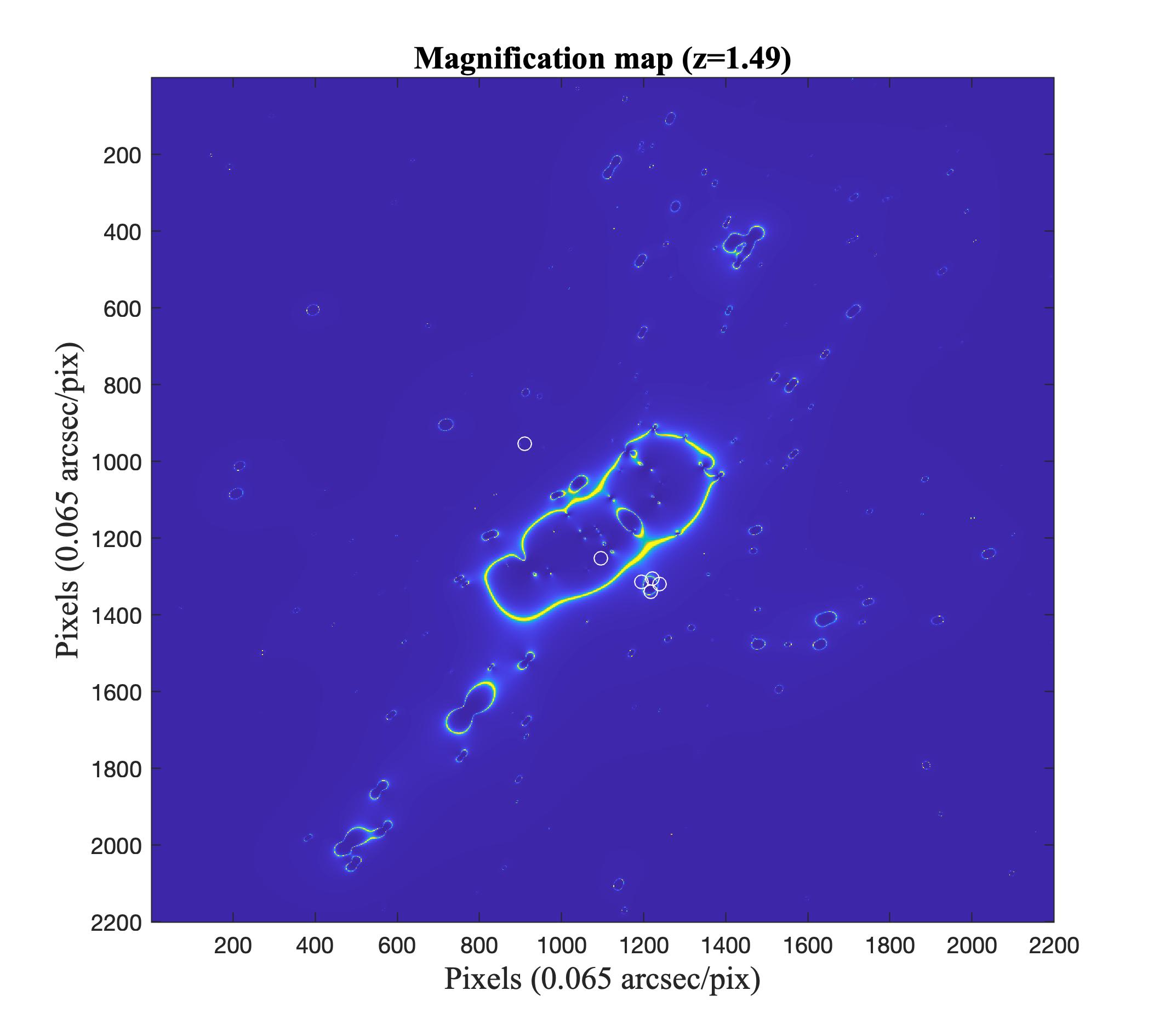}\\
      \includegraphics[width=0.35\textwidth,trim=0cm 0cm 0cm 0cm,clip]{TDyears.jpg}
      \includegraphics[width=0.35\textwidth,trim=0cm 0cm 0cm 0cm,clip]{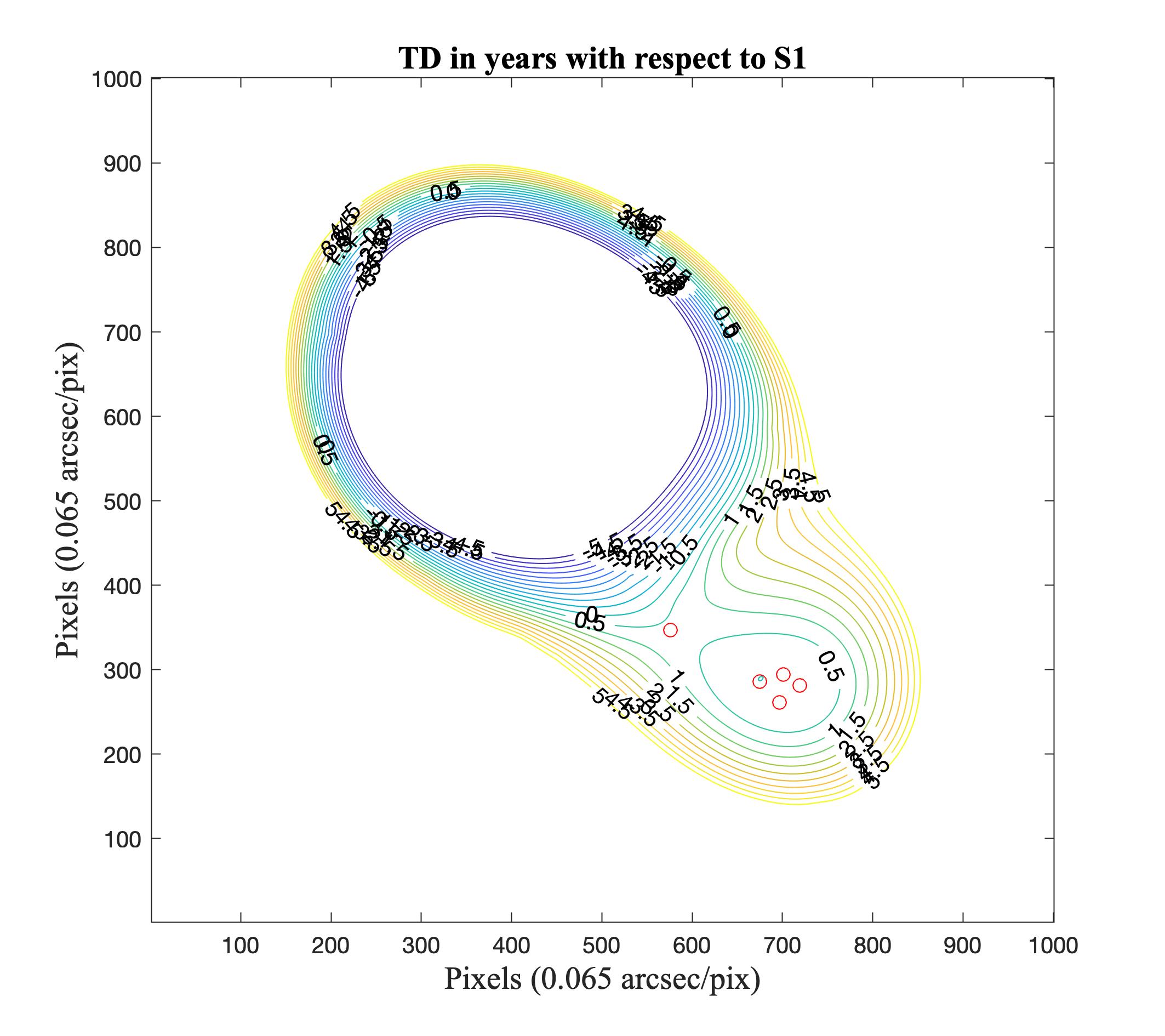}\\
    \includegraphics[width=0.35\textwidth,trim=0cm 0cm 0cm 0cm,clip]{TDdays.jpg}
    \includegraphics[width=0.35\textwidth,trim=0cm 0cm 0cm 0cm,clip]{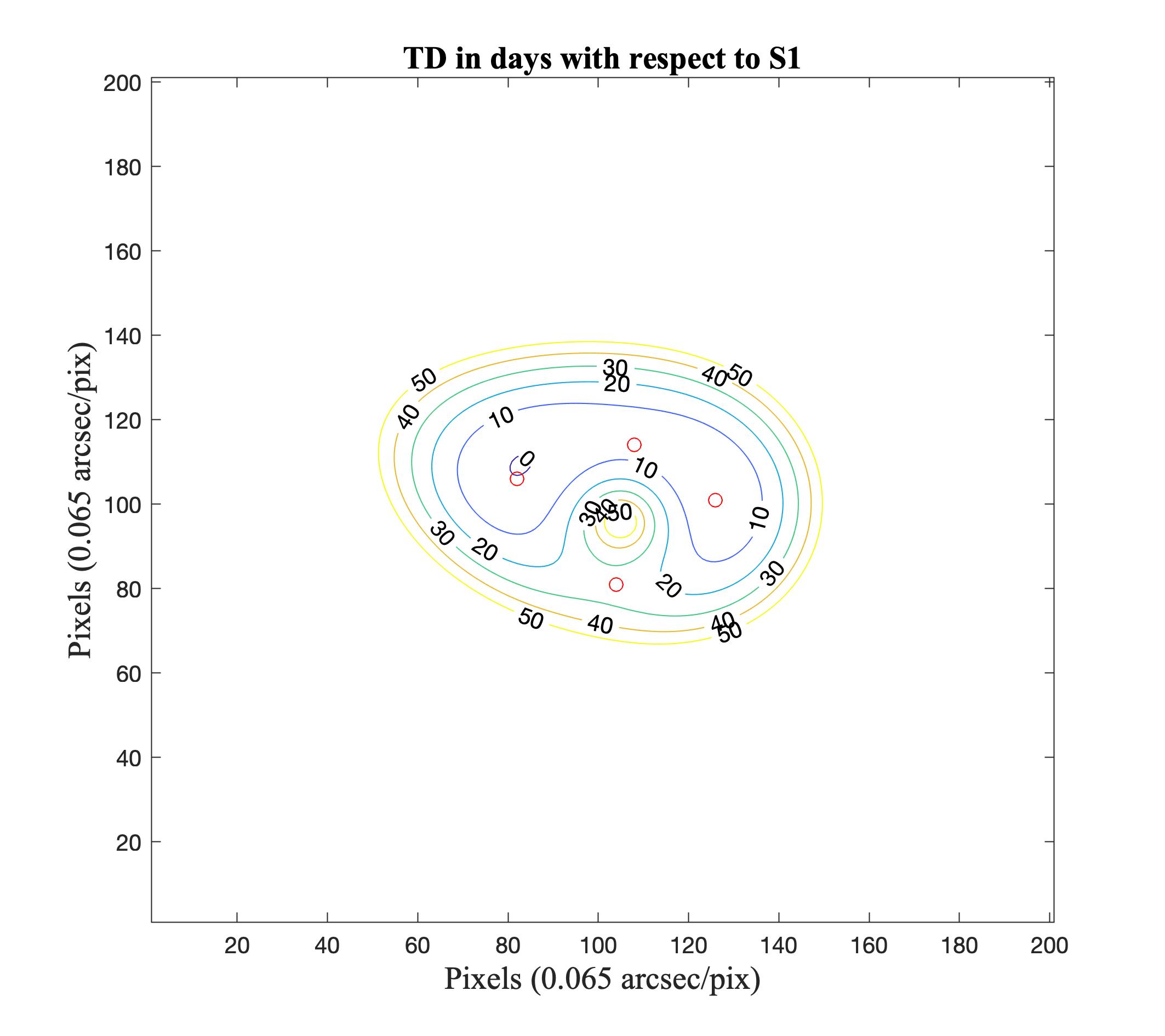}
 \end{center}
\caption{The parametric dPIEeNFW model (\emph{left}) compared with the LTM model for M1149 (\emph{right}). The rows from top to bottom show $\kappa$, the mass density mass; the magnification map; the TD surface with respect to S1, in years; and the TD surface, in days, zoomed in on the Einstein cross. See text for discussion.}\label{fig:compwithLTM}
\end{figure*}

In Fig. \ref{fig:compwithLTM} we show again the properties of our dPIEeNFW model, side-by-side with those of our LTM model. Both models were constructed using the same pipeline, with the same grid resolution, same list of cluster members as input and same multiple image constraints. The main difference between the two methods lies in the parametrization. While it is unclear at present which exact properties of the parametrizations play a significant role here (this naturally deserves a separate and more quantitative examination, to be done elsewhere), we can speculate on the potential causes. In particular, we list three principal differences between the two parametrizations, and discuss to what extent they may affect the solution:
\smallskip

\noindent(i). Cluster galaxies are generally modeled as dPIE in the parametric model, i.e., they have a pseudo isothermal mass density profile and a cut-off radius. In contrast, galaxies in the LTM methodology are modeled each as a power-law mass distribution, with no cut-off, and a profile steeper than isothermal (also, the power-law exponent is typically a free parameter because it affects the overall profile of the cluster). In both methodologies galaxies are modeled as circular with only a few key ones having ellipticity assigned to them. 

On one hand, given most of the lensing occurs on cluster scales, and since the deflection angle from each mass component only depends on the interior mass within some radius of interest, the exact small-scale distribution should not have a major effect on the global model properties. On the other hand, since the Einstein-cross images are primarily lensed by a cluster \emph{galaxy}, smaller-scale perturbation by this galaxy could be important, so the different parametrization may affect the solutions for the cross. Nevertheless, the two TD surfaces concentrated on this galaxy, shown in the bottom row of Fig. \ref{fig:compwithLTM}, do not seem to reveal a significant difference, indicating that the exact parametrization also of this galaxy, potentially, does not play a crucial role in e.g., the SX-S1 TD. It could account, however, for the more minor (but statistically significant) differences between the Einstein cross images. 

Note also, the uncertainties on the magnification ratios from our parametric model are often significantly smaller than those from other parametric lens models. This could be a result of other models independently modeling the Einstein cross lens galaxy, whereas in our case, besides its mass, we left its properties coupled to the scaling relations.
\smallskip

\noindent(ii). The dark matter distribution in the parametric case is modeled nearly-independently of the galaxy component, or the galaxy light distribution, whereas in the LTM methodology the dark matter map is coupled to the galaxy component (the dark matter map is a smoothed version of the galaxies' luminosity-weighted distribution). In that sense the parametric method has more freedom and flexibility to describe the multiple images more closely, and often (including here) results in an overall better fit to the data, and thus, in principle, higher accuracy. On the other hand, the LTM can probe a different range of DM profile shapes as it is not coupled to a certain analytical form (or combinations thereof), and it has unprecedented prediction power to predict the appearance of multiple images based solely on the luminosity distribution as input \citep[e.g.][]{Carrasco2020,Zalesky2020}. It is tempting to claim that the somewhat higher accuracy of the dPIEeNFW model ($\emph{rms}=0.57\arcsec$) compared to the LTM model ($\emph{rms}=0.68\arcsec$)  leads to better estimates for SN Refsdal, and SX-S1 in particular. However, the LTM model has half the number of free parameters, and it is conceivable that future versions, allowing for more free parameters (more free galaxy masses, for example), will help improve the fit further. In addition, given that the predictions from the range of probed LTM models constructed in the process of this work, were systematically lower than the range from the parametric models constructed, the higher accuracy of the parametric model does not necessarily seem to be the cause.
\smallskip

\noindent(iii). Another difference that follows the different dark matter representation in the two methods is the role, or implementation, of ellipticity. The parametric model we employ here includes elliptical NFW halos for the dark matter, so that the ellipticity of each halo is embedded in the mass distribution itself, whereas in the LTM model there is no global ellipticity embedded in the mass distribution (only that introduced for individual galaxies). The parametric model also does not require an external shear to obtain a good fit, whereas the external shear in the LTM case was found to be quite strong (about $\gamma_{ex}\simeq0.2$ along the main elongation direction of the cluster), imitating ellipticity in the critical curves, but the mass distribution itself is evidently rounder (Fig. \ref{fig:compwithLTM}). In addition, unlike the LTM model, the parametric modeling has been shown to require a third, bright or massive clump north of the BCG ($\sim1.2\times10^{14}$ M$_{\odot}$ in our model; see also Fig. \ref{fig:compwithLTM}). This bright clump may partly replace the need for an external shear, although it produces different effects in the mass distribution (and critical curves). The lack of constraints in the northern part of the lens around that third clump does not allow for a detailed determination of its true weight. Alternatively, the lack of images may suggest, unless contributed by local, cosmic variance, that perhaps that third clump is not as massive as indicated by most parametric methods, and thus may hint that the LTM parametrization is more correct around that clump, in that sense. Future data (as well as other measurements such as  weak lensing, possibly) will be needed to examine the weight of that clump and the possible degeneracies between this clump, the inherent matter ellipticity, and an external shear. 

More detailed examinations of these features are warranted, and remain for future work.
\smallskip

The effect of the adopted source position on the TDs should also be discussed (e.g., \citealt{BirrerTreu2019}). For our LTM model in Paper I, the TDs were affected by the exact position of the source, which can be obtained in various ways. We considered two source positions to derive the TDs, and the SX-S1 TD in particular. One source position was obtained by delensing only the four images of the Einstein cross and taking their average source position. The second, averaged source position included also the delensed position of SX in the average. The second source position yielded SX-S1 LTM TDs that are about 40 - 50 days larger than with the first source position (this higher TD likely more correct; although here we adopt the former for a fair comparison), and it changed the TDs of the Einstein cross images by order days. For our dPIEeNFW model we find, perhaps due to the higher accuracy, that the SX-S1 TD increases by about 15 days only, if considering the second source position, compared to the nominal values listed in Table \ref{TDs} using the first source position. The effect on the Einstein cross is smaller, of the order of days, similar to what was seen in the LTM case.

It may be somewhat surprising that while the two mass maps in Fig. \ref{fig:compwithLTM} are notably, quite distinct (upper row; especially, the LTM model seems to have more mass associated with bright galaxies also away for the center, compared to the more concentrated dPIEeNFW model), the overall reproduction \emph{rms} is not very different (15\% better for the parametric method). In addition, the overall TD surface (third row) seems to be relatively similar in shape, but with some subtle differences near the position of SX, leading to shorter delays in the flatter, LTM case.

\subsection{Implications for the Hubble constant}

The equation that describes the delayed arrival time of each image, due to the presence of the lens, compared to an undeflected light ray from the same source, is given by (\citealt{NarayanBartelmann1996}):
\begin{equation}\label{eq:TD}
    t(\vec\theta)=\frac{1+z_{l}}{c}\frac{D_{l}D_{s}}{D_{ls}}\left[\frac{1}{2}(\vec\theta-\vec\beta)^2-\psi({\vec\theta})\right]
\end{equation}
where $\vec\theta$ is the image position, $\vec\beta$ is the source position as indicated by the lens model, and $\psi$ the gravitational potential given by the lens model. $D_{l}$, $D_{s}$, and $D_{ls}$ are the angular diameter distances to the lens, to the source, and between the lens and the source, respectively. We denote the arrival time difference between the different multiple images simply as the TD. 

The TD distance, i.e., the combination of angular diameter distances, $\frac{D_{l}D_{s}}{D_{ls}}$ (times ($1+z_{l}$)), is inversely proportional to the Hubble constant. Thus, by comparing measured TDs with those predicted from a lens model, the Hubble constant can be constrained. To a reaosnable approximation, one can obtain a rough estimate by simply rescaling the model's TD, using the measured TD: 
\begin{equation}\label{eq:TD2}
H_{0,true}=\text{TD}_{model}/\text{TD}_{true} \cdot H_{0,model} 
\end{equation}
where we used in our modeling $H_{0,model}=70$ km/s/Mpc. The final, measured TD for Refsdal is yet unknown at the time of writing. If, for example, we use the TD$_{true}=340^{+43}_{-52}$ between S1 and SX recently measured by \citet{Baklanov2020} (see Fig. \ref{fig:compwithMeasurements} here), the Hubble constant implied by our parametric model would be $\simeq67\pm10$ km/s/Mpc. The measurement in \citet{Kelly2016reappearance} is similar, concentrated around $\sim345$ days (with about a 10\% error). For comparison, as our LTM model suggests systematically smaller SX-S1 TDs, by about 20-30\%, it correspondingly leads to a smaller Hubble constant, for the same measurement by \citet{Baklanov2020} or \citet{Kelly2016reappearance}, although the LTM prediction likely underestimates the true reappearance time. 

An another comparison, note that previous measurements of the Hubble constant with SN Refsdal took place before (e.g., see \citealt{Vega-Ferrero2018RefsdalH0} and \citealt{Grillo2018H0m1149}, \citealt{Baklanov2020}), that are generally, and unsurprisingly, in good agreement with our measurement. Possible systematics and the accuracy of the TDs and the constraints on the Hubble constant were investigated, for example, by \citet{Williams2019TDaccuracy1149} and \citet{Grillo2020TDaccuracy1149}. A more accurate determination of the value for the Hubble constant and obtained from an ensemble of models could be made when more accurate measurements of the observed TDs and relative magnifications become available, e.g., exploiting the longer monitoring of SN Refsdal compared to the data in \citet{Kelly2016reappearance}.

\section{Summary}\label{s:summary}
The main purpose of this paper was to present a new, grid-based parametric model for M1149, concentrating on the properties of SN Refsdal. The model complements our LTM model for M1149 published in Paper I, and was built on the same grid, using the same pipeline and with similar inputs, so that a comparison between the two parametrizations is facilitated. No TD or magnification ratio information was used in the minimization.%, and we anticipate that more accurate measurements for SN Refsdal's properties will become available in the near future. 

We compare our estimates with inferred properties of SN Refsdal and with predictions from other parametric lens models submitted to SN Refsdal's blind prediction challenge as published in \citet[][]{Treu2016Refsdal} and \citet{Kelly2016reappearance}, and references therein. The TDs and magnification ratios we obtain for SN Refsdal seem to be in very good agreement (to within 1-2$\sigma$ typically) with those from most other parametric models, as well as with the early measurements by \citet{Kelly2016reappearance,Rodney2016Refsdal,Treu2016Refsdal} and with the recent measurements by \citet{Baklanov2020}. Adopting the measurements for the SX-S1 TD by the latter, the Hubble constant implied by our new parametric model is $\simeq67\pm10$ km/s/Mpc, where the error encompasses the typical range of values from our parametric trial models. %This value should be revised and improved when refined measurements of the TDs and magnification ratios become available.

We also examine here differences between our recent LTM model (Paper I) and our dPIEeNFW model. We compare their TD predictions for SN Refsdal, finding that the predictions from the parametric method are in better agreement with those from most other parametric methods (and especially the models by \emph{Grillo} and \emph{Oguri}) but can differ more substantially from the LTM estimates. In addition, we find that the magnification ratios implied by the LTM model, with respect to S1, are generally smaller compared to the parametric technique, and that for S2, S3, and SX, the TDs with respect to S1 are systematically lower in the LTM model compared to the parametric model. We compare side-by-side the mass distributions, magnification maps, and TD surfaces from the two models, and speculate on factors that may lead to the different TD estimates, such as role of individual galaxies and their representation, the representation of the dark-matter component and its internal ellipticity versus external shear or a massive substructure, and whether the accuracy of the model is necessarily an indicator for the correctness of its TD prediction. Updated measurements of SN Refsdal could help decipher between the two solutions, although it seems that -- most notably for SX-S1 -- the parametric models give more accurate estimates, and the LTM model likely underestimates the true TD.

Most lensing models rely solely on the position of multiple images as constraints. Refined TD and magnification ratio measurements for SN Refsdal should allow to incorporate additional information in the lens models, and break some of the degeneracies inherent to common lensing analyses, and thus, potentially, improve the constraints on the underlying matter distribution of the lens, including the intrinsic shape of the unseen, dark matter component.

\smallskip

\section*{acknowledgements}
I kindly thank Pat Kelly, Tommaso Treu, and Steve Rodney, for useful discussions. The work uses some scripts from the astronomy \textsc{Matlab} package \citep{OfekMatlab2014ascl.soft07005O}. I am also grateful for useful discussions and multiple image voting that took place as a community effort in the HFF framework, and to the respective teams that submitted models  -- led by PIs Bradac, Natarajan \& Kneib (CATS), Merten \& Zitrin, Sharon, Williams, Keeton, Bernstein and Diego, and the GLAFIC group. This work is based on observations obtained with the NASA/ESA Hubble Space Telescope, retrieved from the Mikulski Archive for Space Telescopes (MAST) at the Space Telescope Science Institute (STScI). STScI is operated by the Association of Universities for Research in Astronomy, Inc. under NASA contract NAS 5-26555.


\begin{thebibliography}{}
\expandafter\ifx\csname natexlab\endcsname\relax\def\natexlab#1{#1}\fi
\providecommand{\url}[1]{\href{#1}{#1}}

\bibitem[{{Acebron} {et~al.}(2017){Acebron}, {Jullo}, {Limousin}, {Tilquin},
  {Giocoli}, {Jauzac}, {Mahler}, \& {Richard}}]{Acebron2017Systematics}
{Acebron}, A., {Jullo}, E., {Limousin}, M., {et~al.} 2017, \mnras, 470, 1809

\bibitem[{{Acebron} {et~al.}(2019){Acebron}, {Zitrin}, {Coe}, {Mahler},
  {Sharon}, {Oguri}, {Brada{\v{c}}}, {Bradley}, {Frye}, {Forman}, {Strait},
  {Su}, {Umetsu}, {Andrade-Santos}, {Avila}, {Carrasco}, {Cerny}, {Czakon},
  {Dawson}, {Fox}, {Hoag}, {Huang}, {Johnson}, {Kikuchihara}, {Lam},
  {Lovisari}, {Mainali}, {Nonino}, {Oesch}, {Ogaz}, {Ouchi}, {Past},
  {Paterno-Mahler}, {Peterson}, {Ryan}, {Salmon}, {Stark}, {Toft}, {Trenti},
  {Vulcani}, \& {Welch}}]{Acebron2019RXC0032}
{Acebron}, A., {Zitrin}, A., {Coe}, D., {et~al.} 2019, arXiv e-prints,
  arXiv:1912.02702

\bibitem[{{Atek} {et~al.}(2015){Atek}, {Richard}, {Jauzac}, {Kneib},
  {Natarajan}, {Limousin}, {Schaerer}, {Jullo}, {Ebeling}, {Egami}, \&
  {Clement}}]{Atek2015HalfHFFLF}
{Atek}, H., {Richard}, J., {Jauzac}, M., {et~al.} 2015, \apj, 814, 69

\bibitem[{{Baklanov} {et~al.}(2020){Baklanov}, {Lyskova}, {Blinnikov}, \&
  {Nomoto}}]{Baklanov2020}
{Baklanov}, P., {Lyskova}, N., {Blinnikov}, S., \& {Nomoto}, K. 2020, arXiv
  e-prints, arXiv:2007.04106

\bibitem[{{Birrer} \& {Treu}(2019)}]{BirrerTreu2019}
{Birrer}, S., \& {Treu}, T. 2019, \mnras, 489, 2097

\bibitem[{{Bolton} {et~al.}(2006){Bolton}, {Burles}, {Koopmans}, {Treu}, \&
  {Moustakas}}]{Bolton2006SLACS}
{Bolton}, A.~S., {Burles}, S., {Koopmans}, L. V.~E., {Treu}, T., \&
  {Moustakas}, L.~A. 2006, \apj, 638, 703

\bibitem[{{Brada{\v c}} {et~al.}(2012){Brada{\v c}}, {Vanzella}, {Hall},
  {Treu}, {Fontana}, {Gonzalez}, {Clowe}, {Zaritsky}, {Stiavelli}, \&
  {Cl{\'e}ment}}]{Bradac2012highz}
{Brada{\v c}}, M., {Vanzella}, E., {Hall}, N., {et~al.} 2012, \apjl, 755, L7

\bibitem[{{Broadhurst} {et~al.}(2005){Broadhurst}, {Ben{\'{\i}}tez}, {Coe},
  {Sharon}, {Zekser}, {White}, {Ford}, {Bouwens}, {Blakeslee}, {Clampin},
  {et~al.}}]{Broadhurst2005a}
{Broadhurst}, T., {Ben{\'{\i}}tez}, N., {Coe}, D., {et~al.} 2005, \apj, 621, 53

\bibitem[{{Caminha} {et~al.}(2016){Caminha}, {Grillo}, {Rosati}, {Balestra},
  {Karman}, {Lombardi}, {Mercurio}, {Nonino}, {Tozzi}, {Zitrin}, {Biviano},
  {Girardi}, {Koekemoer}, {Melchior}, {Meneghetti}, {Munari}, {Suyu}, {Umetsu},
  {Annunziatella}, {Borgani}, {Broadhurst}, {Caputi}, {Coe}, {Delgado-Correal},
  {Ettori}, {Fritz}, {Frye}, {Gobat}, {Maier}, {Monna}, {Postman}, {Sartoris},
  {Seitz}, {Vanzella}, \& {Ziegler}}]{Caminha2016RXJ}
{Caminha}, G.~B., {Grillo}, C., {Rosati}, P., {et~al.} 2016, \aap, 587, A80

\bibitem[{{Carrasco} {et~al.}(2020){Carrasco}, {Zitrin}, \&
  {Seidel}}]{Carrasco2020}
{Carrasco}, M., {Zitrin}, A., \& {Seidel}, G. 2020, \mnras, 491, 3778

\bibitem[{{Cerny} {et~al.}(2018){Cerny}, {Sharon}, {Andrade-Santos}, {Avila},
  {Brada{\v{c}}}, {Bradley}, {Carrasco}, {Coe}, {Czakon}, {Dawson}, {Frye},
  {Hoag}, {Huang}, {Johnson}, {Jones}, {Lam}, {Lovisari}, {Mainali}, {Oesch},
  {Ogaz}, {Past}, {Paterno-Mahler}, {Peterson}, {Riess}, {Rodney}, {Ryan},
  {Salmon}, {Sendra-Server}, {Stark}, {Strolger}, {Trenti}, {Umetsu},
  {Vulcani}, \& {Zitrin}}]{Cerny2018}
{Cerny}, C., {Sharon}, K., {Andrade-Santos}, F., {et~al.} 2018, \apj, 859, 159

\bibitem[{{Chen} {et~al.}(2020){Chen}, {Broadhurst}, {Lim}, {Molnar}, {Diego},
  {Oguri}, \& {Lee}}]{Chen2020A3827}
{Chen}, M.~C., {Broadhurst}, T., {Lim}, J., {et~al.} 2020, arXiv e-prints,
  arXiv:2007.05603

\bibitem[{{Coe} {et~al.}(2015){Coe}, {Bradley}, \& {Zitrin}}]{Coe2014FF}
{Coe}, D., {Bradley}, L., \& {Zitrin}, A. 2015, \apj, 800, 84

\bibitem[{{Courbin} {et~al.}(2011){Courbin}, {Chantry}, {Revaz}, {Sluse},
  {Faure}, {Tewes}, {Eulaers}, {Koleva}, {Asfandiyarov}, {Dye}, {Magain}, {van
  Winckel}, {Coles}, {Saha}, {Ibrahimov}, \& {Meylan}}]{Courbin2011}
{Courbin}, F., {Chantry}, V., {Revaz}, Y., {et~al.} 2011, \aap, 536, A53

\bibitem[{{Diego} {et~al.}(2005){Diego}, {Protopapas}, {Sandvik}, \&
  {Tegmark}}]{Diego2005Nonparam}
{Diego}, J.~M., {Protopapas}, P., {Sandvik}, H.~B., \& {Tegmark}, M. 2005,
  \mnras, 360, 477

\bibitem[{{Diego} {et~al.}(2016){Diego}, {Broadhurst}, {Chen}, {Lim}, {Zitrin},
  {Chan}, {Coe}, {Ford}, {Lam}, \& {Zheng}}]{Diego2016refsdal}
{Diego}, J.~M., {Broadhurst}, T., {Chen}, C., {et~al.} 2016, \mnras, 456, 356

\bibitem[{{Ebeling} {et~al.}(2007){Ebeling}, {Barrett}, {Donovan}, {Ma},
  {Edge}, \& {van Speybroeck}}]{EbelingMacs12_2007}
{Ebeling}, H., {Barrett}, E., {Donovan}, D., {et~al.} 2007, \apjl, 661, L33

\bibitem[{{El{\'\i}asd{\'o}ttir} {et~al.}(2007){El{\'\i}asd{\'o}ttir},
  {Limousin}, {Richard}, {Hjorth}, {Kneib}, {Natarajan}, {Pedersen}, {Jullo},
  \& {Paraficz}}]{Eliasdottir2007}
{El{\'\i}asd{\'o}ttir}, {\'A}., {Limousin}, M., {Richard}, J., {et~al.} 2007,
  arXiv e-prints, arXiv:0710.5636

\bibitem[{{Finney} {et~al.}(2018){Finney}, {Brada{\v{c}}}, {Huang}, {Hoag},
  {Morishita}, {Schrabback}, {Treu}, {Borello Schmidt}, {Lemaux}, {Wang}, \&
  {Mason}}]{Finney2018M1149}
{Finney}, E.~Q., {Brada{\v{c}}}, M., {Huang}, K.-H., {et~al.} 2018, \apj, 859,
  58

\bibitem[{{Freedman} {et~al.}(2019){Freedman}, {Madore}, {Hatt}, {Hoyt},
  {Jang}, {Beaton}, {Burns}, {Lee}, {Monson}, {Neeley}, {Phillips}, {Rich}, \&
  {Seibert}}]{Freedman2019H0}
{Freedman}, W.~L., {Madore}, B.~F., {Hatt}, D., {et~al.} 2019, \apj, 882, 34

\bibitem[{{Grillo} {et~al.}(2020){Grillo}, {Rosati}, {Suyu}, {Caminha},
  {Mercurio}, \& {Halkola}}]{Grillo2020TDaccuracy1149}
{Grillo}, C., {Rosati}, P., {Suyu}, S.~H., {et~al.} 2020, arXiv e-prints,
  arXiv:2001.02232

\bibitem[{{Grillo} {et~al.}(2018){Grillo}, {Rosati}, {Suyu}, {Balestra},
  {Caminha}, {Halkola}, {Kelly}, {Lombardi}, {Mercurio}, {Rodney}, \&
  {Treu}}]{Grillo2018H0m1149}
---. 2018, \apj, 860, 94

\bibitem[{{Halkola} {et~al.}(2006){Halkola}, {Seitz}, \&
  {Pannella}}]{Halkola2006}
{Halkola}, A., {Seitz}, S., \& {Pannella}, M. 2006, \mnras, 372, 1425

\bibitem[{{Holder} \& {Schechter}(2003)}]{HolderSchechter2003Shear}
{Holder}, G.~P., \& {Schechter}, P.~L. 2003, \apj, 589, 688

\bibitem[{{Ishigaki} {et~al.}(2018){Ishigaki}, {Kawamata}, {Ouchi}, {Oguri},
  {Shimasaku}, \& {Ono}}]{Ishigaki2018HFF}
{Ishigaki}, M., {Kawamata}, R., {Ouchi}, M., {et~al.} 2018, \apj, 854, 73

\bibitem[{{Jauzac} {et~al.}(2015){Jauzac}, {Richard}, {Jullo}, {Cl{\'e}ment},
  {Limousin}, {Kneib}, {Ebeling}, {Natarajan}, {Rodney}, {Atek}, {Massey},
  {Eckert}, {Egami}, \& {Rexroth}}]{Jauzac2015A2744}
{Jauzac}, M., {Richard}, J., {Jullo}, E., {et~al.} 2015, \mnras, 452, 1437

\bibitem[{{Jing} \& {Suto}(2000)}]{JingSuto2000}
{Jing}, Y.~P., \& {Suto}, Y. 2000, \apjl, 529, L69

\bibitem[{{Johnson} \& {Sharon}(2016)}]{Johnson2016systematics}
{Johnson}, T.~L., \& {Sharon}, K. 2016, \apj, 832, 82

\bibitem[{{Jullo} {et~al.}(2007){Jullo}, {Kneib}, {Limousin},
  {El{\'{\i}}asd{\'o}ttir}, {Marshall}, \& {Verdugo}}]{Jullo2007Lenstool}
{Jullo}, E., {Kneib}, J.-P., {Limousin}, M., {et~al.} 2007, New Journal of
  Physics, 9, 447

\bibitem[{{Jullo} {et~al.}(2010){Jullo}, {Natarajan}, {Kneib}, {D'Aloisio},
  {Limousin}, {Richard}, \& {Schimd}}]{Jullo2010}
{Jullo}, E., {Natarajan}, P., {Kneib}, J., {et~al.} 2010, Science, 329, 924

\bibitem[{{Kawamata} {et~al.}(2016){Kawamata}, {Oguri}, {Ishigaki},
  {Shimasaku}, \& {Ouchi}}]{Kawamata2016modelsHFF}
{Kawamata}, R., {Oguri}, M., {Ishigaki}, M., {Shimasaku}, K., \& {Ouchi}, M.
  2016, \apj, 819, 114

\bibitem[{{Keeton} {et~al.}(1997){Keeton}, {Kochanek}, \&
  {Seljak}}]{Keeton1997Shear}
{Keeton}, C.~R., {Kochanek}, C.~S., \& {Seljak}, U. 1997, \apj, 482, 604

\bibitem[{{Kelly} {et~al.}(2015){Kelly}, {Rodney}, {Treu}, {Foley}, {Brammer},
  {Schmidt}, {Zitrin}, {Sonnenfeld}, {Strolger}, {Graur}, {Filippenko}, {Jha},
  {Riess}, {Bradac}, {Weiner}, {Scolnic}, {Malkan}, {von der Linden}, {Trenti},
  {Hjorth}, {Gavazzi}, {Fontana}, {Merten}, {McCully}, {Jones}, {Postman},
  {Dressler}, {Patel}, {Cenko}, {Graham}, \& {Tucker}}]{Kelly2015Sci}
{Kelly}, P.~L., {Rodney}, S.~A., {Treu}, T., {et~al.} 2015, Science, 347, 1123

\bibitem[{{Kelly} {et~al.}(2016){Kelly}, {Rodney}, {Treu}, {Strolger}, {Foley},
  {Jha}, {Selsing}, {Brammer}, {Brada{\v c}}, {Cenko}, {Graur}, {Filippenko},
  {Hjorth}, {McCully}, {Molino}, {Nonino}, {Riess}, {Schmidt}, {Tucker}, {von
  der Linden}, {Weiner}, \& {Zitrin}}]{Kelly2016reappearance}
---. 2016, \apjl, 819, L8

\bibitem[{{Kneib} {et~al.}(1996){Kneib}, {Ellis}, {Smail}, {Couch}, \&
  {Sharples}}]{Kneib1996}
{Kneib}, J.-P., {Ellis}, R.~S., {Smail}, I., {Couch}, W.~J., \& {Sharples},
  R.~M. 1996, \apj, 471, 643

\bibitem[{{Kneib} \& {Natarajan}(2011)}]{Kneib2011review}
{Kneib}, J.-P., \& {Natarajan}, P. 2011, \aapr, 19, 47

\bibitem[{{Kochanek} {et~al.}(2001){Kochanek}, {Keeton}, \&
  {McLeod}}]{Kochanek2001Ring}
{Kochanek}, C.~S., {Keeton}, C.~R., \& {McLeod}, B.~A. 2001, \apj, 547, 50

\bibitem[{{Kovner}(1987)}]{Kovner1987}
{Kovner}, I. 1987, \apj, 312, 22

\bibitem[{{Liesenborgs} {et~al.}(2006){Liesenborgs}, {De Rijcke}, \&
  {Dejonghe}}]{Liesenborgs2006}
{Liesenborgs}, J., {De Rijcke}, S., \& {Dejonghe}, H. 2006, \mnras, 367, 1209

\bibitem[{{Limousin} {et~al.}(2010){Limousin}, {Ebeling}, {Ma}, {Swinbank},
  {Smith}, {Richard}, {Edge}, {Jauzac}, {Kneib}, {Marshall}, \&
  {Schrabback}}]{Limousin2010M1423}
{Limousin}, M., {Ebeling}, H., {Ma}, C.-J., {et~al.} 2010, \mnras, 405, 777

\bibitem[{{Lotz} {et~al.}(2017){Lotz}, {Koekemoer}, {Coe}, {Grogin}, {Capak},
  {Mack}, {Anderson}, {Avila}, {Barker}, {Borncamp}, {Brammer}, {Durbin},
  {Gunning}, {Hilbert}, {Jenkner}, {Khandrika}, {Levay}, {Lucas}, {MacKenty},
  {Ogaz}, {Porterfield}, {Reid}, {Robberto}, {Royle}, {Smith},
  {Storrie-Lombardi}, {Sunnquist}, {Surace}, {Taylor}, {Williams}, {Bullock},
  {Dickinson}, {Finkelstein}, {Natarajan}, {Richard}, {Robertson}, {Tumlinson},
  {Zitrin}, {Flanagan}, {Sembach}, {Soifer}, \& {Mountain}}]{Lotz2017HFF}
{Lotz}, J.~M., {Koekemoer}, A., {Coe}, D., {et~al.} 2017, \apj, 837, 97

\bibitem[{{Massey} {et~al.}(2015){Massey}, {Williams}, {Smit}, {Swinbank},
  {Kitching}, {Harvey}, {Jauzac}, {Israel}, {Clowe}, {Edge}, {Hilton}, {Jullo},
  {Leonard}, {Liesenborgs}, {Merten}, {Mohammed}, {Nagai}, {Richard},
  {Robertson}, {Saha}, {Santana}, {Stott}, \& {Tittley}}]{Massey2015}
{Massey}, R., {Williams}, L., {Smit}, R., {et~al.} 2015, \mnras, 449, 3393

\bibitem[{{McLeod} {et~al.}(2016){McLeod}, {McLure}, \& {Dunlop}}]{Mcleod2016}
{McLeod}, D.~J., {McLure}, R.~J., \& {Dunlop}, J.~S. 2016, \mnras, 459, 3812

\bibitem[{{Meneghetti} {et~al.}(2003){Meneghetti}, {Bartelmann}, \&
  {Moscardini}}]{Meneghetti2003b}
{Meneghetti}, M., {Bartelmann}, M., \& {Moscardini}, L. 2003, \mnras, 340, 105

\bibitem[{{Meneghetti} {et~al.}(2016){Meneghetti}, {Natarajan}, {Coe},
  {Contini}, {De Lucia}, {Giocoli}, {Acebron}, {Borgani}, {Bradac}, {Diego},
  {Hoag}, {Ishigaki}, {Johnson}, {Jullo}, {Kawamata}, {Lam}, {Limousin},
  {Liesenborgs}, {Oguri}, {Sebesta}, {Sharon}, {Williams}, \&
  {Zitrin}}]{Meneghetti2016SIMSCOMP}
{Meneghetti}, M., {Natarajan}, P., {Coe}, D., {et~al.} 2016, arXiv, 1606.04548,
  arXiv:1606.04548

\bibitem[{{Meylan} {et~al.}(2006){Meylan}, {Jetzer}, {North}, {Schneider},
  {Kochanek}, \& {Wambsganss}}]{Meylan2006lensing}
{Meylan}, G., {Jetzer}, P., {North}, P., {et~al.}, eds. 2006, {Gravitational
  Lensing: Strong, Weak and Micro}

\bibitem[{{Monna} {et~al.}(2017){Monna}, {Seitz}, {Geller}, {Zitrin},
  {Mercurio}, {Suyu}, {Postman}, {Fabricant}, {Hwang}, \&
  {Koekemoer}}]{Monna2017a611}
{Monna}, A., {Seitz}, S., {Geller}, M.~J., {et~al.} 2017, \mnras, 465, 4589

\bibitem[{{Narayan} \& {Bartelmann}(1996)}]{NarayanBartelmann1996}
{Narayan}, R., \& {Bartelmann}, M. 1996, ArXiv Astrophysics e-prints,
  astro-ph/9606001

\bibitem[{{Navarro} {et~al.}(1996){Navarro}, {Frenk}, \& {White}}]{Navarro1996}
{Navarro}, J.~F., {Frenk}, C.~S., \& {White}, S.~D.~M. 1996, \apj, 462, 563

\bibitem[{{Oesch} {et~al.}(2015){Oesch}, {Bouwens}, {Illingworth}, {Franx},
  {Ammons}, {van Dokkum}, {Trenti}, \& {Labb{\'e}}}]{Oesch2015shear}
{Oesch}, P.~A., {Bouwens}, R.~J., {Illingworth}, G.~D., {et~al.} 2015, \apj,
  808, 104

\bibitem[{{Ofek}(2014)}]{OfekMatlab2014ascl.soft07005O}
{Ofek}, E.~O. 2014, MATLAB package for astronomy and astrophysics, ASCL Code
  Record, ascl:1407.005

\bibitem[{{Oguri}(2015)}]{Oguri2015Refsdal}
{Oguri}, M. 2015, \mnras, 449, L86

\bibitem[{{Oguri} {et~al.}(2012){Oguri}, {Bayliss}, {Dahle}, {Sharon},
  {Gladders}, {Natarajan}, {Hennawi}, \& {Koester}}]{Oguri201238clusters}
{Oguri}, M., {Bayliss}, M.~B., {Dahle}, H., {et~al.} 2012, \mnras, 420, 3213

\bibitem[{{Oke} \& {Gunn}(1983)}]{Oke1983ABandStandards}
{Oke}, J.~B., \& {Gunn}, J.~E. 1983, \apj, 266, 713

\bibitem[{{Paterno-Mahler} {et~al.}(2018){Paterno-Mahler}, {Sharon}, {Coe},
  {Mahler}, {Cerny}, {Johnson}, {Schrabback}, {Andrade-Santos}, {Avila},
  {Brada{\v{c}}}, {Bradley}, {Carrasco}, {Czakon}, {Dawson}, {Frye}, {Hoag},
  {Huang}, {Jones}, {Lam}, {Livermore}, {Lovisari}, {Mainali}, {Oesch}, {Ogaz},
  {Past}, {Peterson}, {Ryan}, {Salmon}, {Sendra-Server}, {Stark}, {Umetsu},
  {Vulcani}, \& {Zitrin}}]{Paterno-Mahler2018}
{Paterno-Mahler}, R., {Sharon}, K., {Coe}, D., {et~al.} 2018, \apj, 863, 154

\bibitem[{{Planck Collaboration} {et~al.}(2018){Planck Collaboration},
  {Aghanim}, {Akrami}, {Ashdown}, {Aumont}, {Baccigalupi}, {Ballardini},
  {Banday}, {Barreiro}, {Bartolo}, {Basak}, {Battye}, {Benabed}, {Bernard},
  {Bersanelli}, {Bielewicz}, {Bock}, {Bond}, {Borrill}, {Bouchet}, {Boulanger},
  {Bucher}, {Burigana}, {Butler}, {Calabrese}, {Cardoso}, {Carron},
  {Challinor}, {Chiang}, {Chluba}, {Colombo}, {Combet}, {Contreras}, {Crill},
  {Cuttaia}, {de Bernardis}, {de Zotti}, {Delabrouille}, {Delouis}, {Di
  Valentino}, {Diego}, {Dor{\'e}}, {Douspis}, {Ducout}, {Dupac}, {Dusini},
  {Efstathiou}, {Elsner}, {En{\ss}lin}, {Eriksen}, {Fantaye}, {Farhang},
  {Fergusson}, {Fernandez-Cobos}, {Finelli}, {Forastieri}, {Frailis},
  {Fraisse}, {Franceschi}, {Frolov}, {Galeotta}, {Galli}, {Ganga},
  {G{\'e}nova-Santos}, {Gerbino}, {Ghosh}, {Gonz{\'a}lez-Nuevo}, {G{\'o}rski},
  {Gratton}, {Gruppuso}, {Gudmundsson}, {Hamann}, {Handley}, {Hansen},
  {Herranz}, {Hildebrandt}, {Hivon}, {Huang}, {Jaffe}, {Jones}, {Karakci},
  {Keih{\"a}nen}, {Keskitalo}, {Kiiveri}, {Kim}, {Kisner}, {Knox},
  {Krachmalnicoff}, {Kunz}, {Kurki-Suonio}, {Lagache}, {Lamarre}, {Lasenby},
  {Lattanzi}, {Lawrence}, {Le Jeune}, {Lemos}, {Lesgourgues}, {Levrier},
  {Lewis}, {Liguori}, {Lilje}, {Lilley}, {Lindholm}, {L{\'o}pez-Caniego},
  {Lubin}, {Ma}, {Mac{\'\i}as-P{\'e}rez}, {Maggio}, {Maino}, {Mandolesi},
  {Mangilli}, {Marcos-Caballero}, {Maris}, {Martin}, {Martinelli},
  {Mart{\'\i}nez-Gonz{\'a}lez}, {Matarrese}, {Mauri}, {McEwen}, {Meinhold},
  {Melchiorri}, {Mennella}, {Migliaccio}, {Millea}, {Mitra},
  {Miville-Desch{\^e}nes}, {Molinari}, {Montier}, {Morgante}, {Moss}, {Natoli},
  {N{\o}rgaard-Nielsen}, {Pagano}, {Paoletti}, {Partridge}, {Patanchon},
  {Peiris}, {Perrotta}, {Pettorino}, {Piacentini}, {Polastri}, {Polenta},
  {Puget}, {Rachen}, {Reinecke}, {Remazeilles}, {Renzi}, {Rocha}, {Rosset},
  {Roudier}, {Rubi{\~n}o-Mart{\'\i}n}, {Ruiz-Granados}, {Salvati}, {Sandri},
  {Savelainen}, {Scott}, {Shellard}, {Sirignano}, {Sirri}, {Spencer},
  {Sunyaev}, {Suur-Uski}, {Tauber}, {Tavagnacco}, {Tenti}, {Toffolatti},
  {Tomasi}, {Trombetti}, {Valenziano}, {Valiviita}, {Van Tent}, {Vibert},
  {Vielva}, {Villa}, {Vittorio}, {Wand elt}, {Wehus}, {White}, {White},
  {Zacchei}, \& {Zonca}}]{Planck2018params}
{Planck Collaboration}, {Aghanim}, N., {Akrami}, Y., {et~al.} 2018, arXiv
  e-prints, arXiv:1807.06209

\bibitem[{{Refsdal}(1964)}]{Refsdal1964MNRAS}
{Refsdal}, S. 1964, \mnras, 128, 307

\bibitem[{{Richard} {et~al.}(2010){Richard}, {Smith}, {Kneib}, {Ellis},
  {Sanderson}, {Pei}, {Targett}, {Sand}, {Swinbank}, {Dannerbauer}, {Mazzotta},
  {Limousin}, {Egami}, {Jullo}, {Hamilton-Morris}, \&
  {Moran}}]{Richard2010locuss20}
{Richard}, J., {Smith}, G.~P., {Kneib}, J.-P., {et~al.} 2010, \mnras, 404, 325

\bibitem[{{Riess} {et~al.}(2019){Riess}, {Casertano}, {Yuan}, {Macri}, \&
  {Scolnic}}]{Riess2019}
{Riess}, A.~G., {Casertano}, S., {Yuan}, W., {Macri}, L.~M., \& {Scolnic}, D.
  2019, \apj, 876, 85

\bibitem[{{Rodney} {et~al.}(2015){Rodney}, {Patel}, {Scolnic}, {Foley},
  {Molino}, {Brammer}, {Jauzac}, {Brada{\v{c}}}, {Broadhurst}, {Coe}, {Diego},
  {Graur}, {Hjorth}, {Hoag}, {Jha}, {Johnson}, {Kelly}, {Lam}, {McCully},
  {Medezinski}, {Meneghetti}, {Merten}, {Richard}, {Riess}, {Sharon},
  {Strolger}, {Treu}, {Wang}, {Williams}, \& {Zitrin}}]{Rodney2015A2744SN}
{Rodney}, S.~A., {Patel}, B., {Scolnic}, D., {et~al.} 2015, \apj, 811, 70

\bibitem[{{Rodney} {et~al.}(2016){Rodney}, {Strolger}, {Kelly}, {Brada{\v c}},
  {Brammer}, {Filippenko}, {Foley}, {Graur}, {Hjorth}, {Jha}, {McCully},
  {Molino}, {Riess}, {Schmidt}, {Selsing}, {Sharon}, {Treu}, {Weiner}, \&
  {Zitrin}}]{Rodney2016Refsdal}
{Rodney}, S.~A., {Strolger}, L.-G., {Kelly}, P.~L., {et~al.} 2016, \apj, 820,
  50

\bibitem[{{Salmon} {et~al.}(2020){Salmon}, {Coe}, {Bradley}, {Bouwens},
  {Brada{\v{c}}}, {Huang}, {Oesch}, {Stark}, {Sharon}, {Trenti}, {Avila},
  {Ogaz}, {Andrade-Santos}, {Carrasco}, {Cerny}, {Dawson}, {Frye}, {Hoag},
  {Johnson}, {Jones}, {Lam}, {Lovisari}, {Mainali}, {Past}, {Paterno-Mahler},
  {Peterson}, {Riess}, {Rodney}, {Ryan}, {Sendra-Server}, {Strait}, {Strolger},
  {Umetsu}, {Vulcani}, \& {Zitrin}}]{Salmon2020HighzRelics}
{Salmon}, B., {Coe}, D., {Bradley}, L., {et~al.} 2020, \apj, 889, 189

\bibitem[{{Sharon} \& {Johnson}(2015)}]{Sharon2015Refsdal}
{Sharon}, K., \& {Johnson}, T.~L. 2015, \apjl, 800, L26

\bibitem[{{Smith} {et~al.}(2009){Smith}, {Ebeling}, {Limousin}, {Kneib},
  {Swinbank}, {Ma}, {Jauzac}, {Richard}, {Jullo}, {Sand}, {Edge}, \&
  {Smail}}]{Smith2009M1149}
{Smith}, G.~P., {Ebeling}, H., {Limousin}, M., {et~al.} 2009, \apjl, 707, L163

\bibitem[{{Suyu} {et~al.}(2013){Suyu}, {Auger}, {Hilbert}, {Marshall}, {Tewes},
  {Treu}, {Fassnacht}, {Koopmans}, {Sluse}, {Blandford}, {Courbin}, \&
  {Meylan}}]{Suyu2013measured}
{Suyu}, S.~H., {Auger}, M.~W., {Hilbert}, S., {et~al.} 2013, \apj, 766, 70

\bibitem[{{Tewes} {et~al.}(2013){Tewes}, {Courbin}, \& {Meylan}}]{Tewes2013}
{Tewes}, M., {Courbin}, F., \& {Meylan}, G. 2013, \aap, 553, A120

\bibitem[{{Treu} {et~al.}(2016){Treu}, {Brammer}, {Diego}, {Grillo}, {Kelly},
  {Oguri}, {Rodney}, {Rosati}, {Sharon}, {Zitrin}, {Balestra}, {Brada{\v c}},
  {Broadhurst}, {Caminha}, {Halkola}, {Hoag}, {Ishigaki}, {Johnson}, {Karman},
  {Kawamata}, {Mercurio}, {Schmidt}, {Strolger}, {Suyu}, {Filippenko}, {Foley},
  {Jha}, \& {Patel}}]{Treu2016Refsdal}
{Treu}, T., {Brammer}, G., {Diego}, J.~M., {et~al.} 2016, \apj, 817, 60

\bibitem[{{Vega-Ferrero} {et~al.}(2018){Vega-Ferrero}, {Diego}, {Miranda}, \&
  {Bernstein}}]{Vega-Ferrero2018RefsdalH0}
{Vega-Ferrero}, J., {Diego}, J.~M., {Miranda}, V., \& {Bernstein}, G.~M. 2018,
  \apjl, 853, L31

\bibitem[{{Williams} \& {Liesenborgs}(2019)}]{Williams2019TDaccuracy1149}
{Williams}, L. L.~R., \& {Liesenborgs}, J. 2019, \mnras, 482, 5666

\bibitem[{{Williams} \& {Saha}(2011)}]{Williams2011A3827}
{Williams}, L. L.~R., \& {Saha}, P. 2011, \mnras, 415, 448

\bibitem[{{Wong} {et~al.}(2019){Wong}, {Suyu}, {Chen}, {Rusu}, {Millon},
  {Sluse}, {Bonvin}, {Fassnacht}, {Taubenberger}, {Auger}, {Birrer}, {Chan},
  {Courbin}, {Hilbert}, {Tihhonova}, {Treu}, {Agnello}, {Ding}, {Jee},
  {Komatsu}, {Shajib}, {Sonnenfeld}, {Bland ford}, {Koopmans}, {Marshall}, \&
  {Meylan}}]{Wong2019H0}
{Wong}, K.~C., {Suyu}, S.~H., {Chen}, G. C.~F., {et~al.} 2019, arXiv e-prints,
  arXiv:1907.04869

\bibitem[{{Wright} \& {Brainerd}(2000)}]{WrightBRainerd2000NFW}
{Wright}, C.~O., \& {Brainerd}, T.~G. 2000, \apj, 534, 34

\bibitem[{{Zalesky} \& {Ebeling}(2020)}]{Zalesky2020}
{Zalesky}, L., \& {Ebeling}, H. 2020, \mnras, arXiv:2007.12182

\bibitem[{{Zitrin} \& {Broadhurst}(2009)}]{Zitrin2009_macs11495}
{Zitrin}, A., \& {Broadhurst}, T. 2009, \apjl, 703, L132

\bibitem[{{Zitrin} {et~al.}(2013){Zitrin}, {Meneghetti}, {Umetsu},
  {Broadhurst}, {Bartelmann}, {Bouwens}, {Bradley}, {Carrasco}, {Coe}, {Ford},
  {Kelson}, {Koekemoer}, {Medezinski}, {Moustakas}, {Moustakas}, {Nonino},
  {Postman}, {Rosati}, {Seidel}, {Seitz}, {Sendra}, {Shu}, {Vega}, \&
  {Zheng}}]{Zitrin2013M0416}
{Zitrin}, A., {Meneghetti}, M., {Umetsu}, K., {et~al.} 2013, \apjl, 762, L30

\bibitem[{{Zitrin} {et~al.}(2015){Zitrin}, {Fabris}, {Merten}, {Melchior},
  {Meneghetti}, {Koekemoer}, {Coe}, {Maturi}, {Bartelmann}, {Postman},
  {Umetsu}, {Seidel}, {Sendra}, {Broadhurst}, {Balestra}, {Biviano}, {Grillo},
  {Mercurio}, {Nonino}, {Rosati}, {Bradley}, {Carrasco}, {Donahue}, {Ford},
  {Frye}, \& {Moustakas}}]{Zitrin2014CLASH25}
{Zitrin}, A., {Fabris}, A., {Merten}, J., {et~al.} 2015, \apj, 801, 44

\end{thebibliography}
\end{document}